\documentclass[aps,prl,showpacs,notitlepage,nofootinbib,
superscriptaddress,floatfix,showkeys,onecolumn,longbibliography]{revtex4-1}

\usepackage[table]{xcolor} % For coloring cells
\usepackage{geometry} % Optional, for better page layout in the example
\usepackage{hyperref}
\usepackage{amsmath,amssymb}
\usepackage{float}
\usepackage{microtype}
\usepackage{slashed}
\usepackage{graphicx}
\usepackage{bm}
\usepackage{latexsym}
\usepackage{epsfig}
\usepackage{psfrag}
\usepackage{color}
\usepackage[dvipsnames]{xcolor}
\usepackage{subfigure}
\usepackage[section]{placeins}
\usepackage{array}   
\usepackage{colortbl} 
\usepackage{easyReview}
\usepackage{tcolorbox}
\usepackage{xfrac}
\usepackage{booktabs}
\usepackage{comment}
\usepackage{array} % Per celle personalizzate
\def\e1i{\epsilon_{1\mathrm{i}}}

\allowdisplaybreaks[1]

\definecolor{customgreen}{HTML}{c7efcf}
\definecolor{customblue}{HTML}{9cafb7}

\begin{document}
%%%%%%%%%%%%%%%%%%%%%%%%%%%%%%%%%%%%%%%%%%%%%%%%%%%%%%%%%%%%%%%%%%%%%%%%%%%%%%%

\definecolor{myred}{RGB}{131,0,12}
\title{\textcolor{myred}%{\Large{Graduation Prediction in LaunchPad platform: Pump.fun }}
{\Large{Predicting the success of new crypto-tokens: \\the Pump.fun case}}
}

\author{Giulio Marino}
\affiliation{Dipartimento di Fisica ``E. Fermi", Università di Pisa, Largo Pontecorvo 3, 56127 Pisa, Italy}
\affiliation{INFN Sezione di Pisa, Polo Fibonacci, Largo B. Pontecorvo 3, 56127 Pisa, Italy}

\author{Manuel Naviglio}
\affiliation{Scuola Normale Superiore, Pisa, Italy}

\author{Francesco Tarantelli}
\affiliation{Dipartimento di Matematica, Universit\`a di Bologna, Bologna, Italy.}

\author{Fabrizio Lillo}
\affiliation{Scuola Normale Superiore, Pisa, Italy}
%\affiliation{Dipartimento di Matematica, Universit\`a di Bologna, Bologna, Italy.}

\begin{abstract}
    We study the dynamics of token launched on Pump.fun, a Solana-based launchpad platform, to identify the determinants of the token success. Pump.fun employs a bonding curve mechanism to bootstrap initial liquidity possibly leading to graduation to the on-chain market, which can be seen as a token success. We build predictive models of the probability of graduation conditional on the current amount of Solana locked in the bonding curve and a set of explanatory variables that capture structural and behavioral aspects of the launch process. Conditioning the graduation probability on these variables significantly improves its predictive power, providing insights into early-stage market behavior, speculative and manipulative dynamics, and the informational efficiency of bonding-curve-based token launches.
\end{abstract}

\maketitle

%%%%%%%%%%%%%%%%%%%%%%%%%%%%%%%%%%%%%%%%%%%%%%%%%%%%%%%%%%%%%%%%%%%%%%%%%%%%%%%%

\section{Introduction}
 
The advent of blockchain technology has dramatically lowered the barriers to issuing new crypto-assets.
Today, tokens can be minted cheaply and almost instantaneously through web interfaces or applications, requiring only a small amount of capital to cover transaction and deployment fees.
In many cases, tokens are introduced by protocol teams as a mechanism to bootstrap ecosystems, incentivize users, or raise resources for further development.
Prominent examples include governance tokens such as JTO (Jito), MORPHO (Morpho), AAVE (Aave), COMP (Compound), as well as the native PUMP token of the Pump.fun launchpad on Solana.

Once issued, tokens, regardless of whether they are backed by a substantive underlying project, become tradable on centralized and decentralized exchanges, where liquidity and price formation are endogenously determined by investor demand.
Projects that manage to construct credible narratives or perceived utility often attract larger communities of traders, a process typically reflected in higher trading volumes and prices.
However, in the absence of verifiable fundamentals, early token adoption is rarely driven by intrinsic value in the traditional sense.
Instead, it is largely shaped by coordination dynamics among traders, in a mechanism reminiscent of Keynes’ ``beauty contest'' \cite{keynes1936general, allen2006beauty, morris2002social}, where agents attempt to anticipate others’ demand rather than to assess long-run economic worth.

This technological ease of token creation has a direct statistical consequence: crypto markets are characterized by an extreme proliferation of new assets and by a vanishingly small fraction of ``successful'' launches.
On most platforms, the overwhelming majority of newly issued tokens exhibit short lifetimes, low liquidity, or outright malicious behavior.
Empirical evidence from Uniswap-style listings shows that most tokens fail to sustain meaningful trading activity, often due to rug-pulls or contract-level manipulation.
For instance, in~\cite{naviglio2025pricemanipulationschemesnew} we document that approximately $88\%$ of tokens launched on Uniswap v2 in late 2024 can be classified as manipulation schemes.

A crucial source of heterogeneity in these markets arises from the \emph{design of smart contracts}.
The contractual rules governing token creation, transfers, minting, burning, and trading permissions can profoundly affect a token lifecycle independently of market demand.
In particular, certain design choices embed manipulation mechanisms directly into the contract logic, for instance by restricting sell permissions or introducing asymmetric transfer rules.
Such designs, commonly referred to as \emph{honeypots}, mechanically trap liquidity and make apparent early success misleading, as the eventual collapse of the token may be encoded in the contract itself rather than driven by endogenous market forces~\cite{Qin2021QuantifyingBE, TorresHoney, Sun2024SoKCA}.

This intertwining of economic behavior and contract design complicates the identification of meaningful predictors of success in decentralized token markets.
A central empirical challenge is therefore to determine which information, observable \emph{in real time} during a token’s early life, is predictive of its future outcome.
This task is difficult for two fundamental reasons.
First, the base rate of success is extremely low, making ex-ante discrimination statistically challenging.
Second, even when manipulation is not the focus, the very definition of ``success'' is often ambiguous, as it typically requires fixing arbitrary time horizons or price thresholds.

A robust empirical analysis thus requires both (i) a controlled environment that limits heterogeneity unrelated to market dynamics and (ii) a non-arbitrary success criterion that is endogenous to the platform mechanism.

\medskip
\noindent\textbf{Pump.fun as a natural laboratory.}
In this work we exploit Pump.fun~\cite{pumpfun}, a Solana-based launchpad whose design satisfies both requirements and provides an unusually clean stage to study the lifecycle of speculative crypto-assets.
Pump.fun enables the creation of a very large number of tokens every day, generating a high-throughput environment well suited for statistical analysis.
More importantly, the launch mechanism is fully standardized: all tokens are created under the same smart contract, share identical initial conditions, and trade along the same bonding-curve design.
This avoids weird contract-specific behavior, like honeypots or transfer limits, that can skew results on DEXs.

Beyond contract-level homogeneity, Pump.fun also embeds an endogenous and protocol-level notion of early success.
Each token begins trading on a virtual constant-product bonding curve that governs price formation and liquidity accumulation.
Graduation occurs deterministically once cumulative buy pressure pushes the bonding curve beyond a fixed threshold of bonded Solana, triggering migration to a real on-chain AMM pool.
During our sample period, this threshold corresponds to approximately $85$~SOL raised and a market capitalization of about \$69{,}000.
Reaching this threshold does not guarantee long-run viability or economic value, but it certifies that the token has attracted sufficient coordinated demand to complete the launchpad phase.
Importantly, only a very small fraction of tokens reach graduation (about $0.63\%$ in our dataset), producing a sharp empirical separation between failures and successes.

From a methodological perspective, Pump.fun therefore functions as a large-scale experimental platform for observing the birth, evolution, and early termination of speculative assets. 
Because most launches are not tied to informative fundamentals, outcomes are primarily driven by trader behavior, coordination, and attention dynamics, all of which are fully observable on-chain. On top of that, this ecosystem is mostly driven by meme tokens that are not backed by solid projects, where the creator main goal is often to make a quick return from the token they launched. That said, some tokens do try to raise funds to build something real, as shown by initiatives like Pump.fun new investment arm, Pump Fund, which aims to support projects. \cite{dubey2026_pumpfun_pumpfund}.
The transparency of blockchain data further allows us to track individual trader addresses and to construct predictive signals based on trader composition and behavior, an approach that is rarely feasible in traditional financial markets.

These features make Pump.fun an ideal setting to investigate a central question in high-churn token markets: given the observable state of early trading activity, which signals are informative about a token likelihood of success?

\medskip
\noindent\textbf{The contribution of this work.}
This paper provides a systematic and fully on-chain empirical analysis of early-stage token launches on bonding-curve-based platforms, using Pump.fun as a canonical and controlled case study.
Our main contribution is to frame the notion of early success in launchpad environments as a probabilistic object that can be monitored and updated in real time, based exclusively on information observable during a token launch phase.

Methodologically, we depart from project-specific or fundamental evaluations and instead treat graduation as a binary, protocol-defined outcome.
This allows us to construct a scalable framework in which the probability of success is estimated conditionally on the evolving state of the bonding curve and on a set of behavioral and microstructural variables.
The resulting graduation-probability curves provide an interpretable, time-consistent summary of how early trading dynamics shape outcomes in an environment characterized by extreme churn and very low base rates of success.

A first contribution is the identification of the amount of Solana accumulated in the bonding curve as a natural state variable summarizing collective demand.
We show that the conditional probability of graduation given the current bonding-curve state is monotone and converges smoothly to unity near the deterministic threshold.
This baseline curve serves as a real-time gauge of launch progress and establishes a natural breakeven against which additional information can be evaluated.

Second, exploiting the transparency of blockchain data, we go beyond aggregate liquidity measures and study how the \emph{composition} of market participants affects graduation likelihood.
We introduce conditioning variables that capture (i) the intensity of bot-like or algorithmic trading activity, (ii) the speed at which liquidity is accumulated, measured by the number of trades required to reach a given bonding-curve level, (iii) the early participation of historically successful traders, and (iv) the identity and behavior of prolific token creators.
All conditioning variables are constructed in a strictly time-consistent manner, using only information available up to the current point on the bonding curve, ensuring a genuinely predictive interpretation.

Our empirical results highlight that not all paths to the same bonding-curve state are equivalent.
Fast accumulation of liquidity through a small number of trades is the strongest predictor of graduation, dominating other variables across the entire range of \textit{vSol}.
In contrast, markets dominated by bot-like activity exhibit systematically lower graduation probabilities beyond intermediate stages, suggesting that high turnover and algorithmic trading do not translate into sustained capital commitment.
The presence of historically successful traders provides at most a modest and non-monotonic effect, reflecting the dual role of such agents in accelerating early discovery while often engaging in rapid exit strategies.

Third, we introduce an explicit economic breakeven that maps graduation probabilities into a minimal profitability condition under a deliberately naive buy-and-hold strategy.
Although not intended as a realistic trading rule, this breakeven provides an interpretable scale to assess whether the estimated probabilities are economically meaningful given the mechanical price dynamics imposed by the bonding curve.
By comparing conditional probability curves to this breakeven, we clarify which features of early trading behavior are not only statistically predictive, but also relevant from an economic perspective.

Finally, we document and rationalize systematic pump-and-dump behavior by token creators, showing how the transition from virtual to real liquidity at graduation induces incentives for pre-graduation liquidation.
This mechanism-level analysis highlights how design choices embedded in bonding-curve launchpads can shape strategic behavior and redistribute value, independently of project quality.

\section{The Solana blockchain}

Solana is a high-performance public blockchain designed to support fast and cheap transactions at large scale.  
Although conceptually similar to Ethereum in that it enables programmable smart contracts and decentralized applications, its underlying architecture differs substantially in both consensus design and system optimization.

The key innovation of Solana lies in its consensus mechanism, known as \textit{Proof of History} (PoH).  
Unlike traditional blockchains such as Ethereum, which rely purely on Proof of Stake (PoS) or Proof of Work (PoW) to agree on the ordering of transactions, Solana introduces a verifiable cryptographic clock that provides a built-in notion of time.  
Each validator maintains a continuous sequence of hash computations, where every hash depends on the previous one, forming a chain that proves the passage of time between events.  
This allows the network to pre-order transactions before they are broadcast and validated, dramatically reducing coordination overhead among validators.

In practice, Proof of History operates in conjunction with a delegated Proof of Stake (dPoS) consensus layer.  
Validators periodically produce blocks based on the PoH sequence, and their ordering can be verified without waiting for network-wide communication.  
This hybrid design enables extremely high throughput (up to several tens of thousands of transactions per second) and sub-second block times, while maintaining decentralized validation and security guarantees.

In contrast, Ethereum, despite its transition to PoS in 2022, relies on a global consensus round for every block, requiring validators to agree on both block contents and ordering.  
This process ensures strong consistency but limits scalability, with average block times of around 12 seconds and significantly lower throughput.

The architectural differences between Solana and Ethereum have major implications for application design.  
Solana’s high-frequency block production and low latency make it particularly suitable for use cases involving rapid transaction sequences, micro-trades, and real-time state updates.  
These features enable the emergence of platforms such as \textit{Pump.fun}, where token creation, bonding curve evolution, and trading occur in seconds, driven by a continuous stream of on-chain interactions.

In the following section, we introduce the Pump.fun platform and describe how it leverages Solana’s unique performance characteristics to implement a fully on-chain launchpad mechanism based on bonding curves.

\section{The Pump.fun platform}

Pump.fun\footnote{See the official documentation at \url{https://pump.fun}.} is a fully on-chain token launchpad built on the Solana blockchain, designed to automate the process of token creation, price discovery, and early trading through bonding-curve dynamics.
It provides a standardized framework in which any user can deploy a new token at minimal cost (of the order of a few dollars), while initial price formation and liquidity provisioning are entirely managed by smart contracts.
The platform design supports the simultaneous creation and trading of thousands of independent tokens, leveraging Solana’s high throughput and low latency for fast execution.

At its core, Pump.fun consists of two distinct and sequential automated market maker (AMM) environments, both based on a constant-product pricing rule in the spirit of Uniswap-like designs~\cite{adams2021uniswap, Angeris2020CFMM}.
These environments differ in their economic interpretation and reserve structure, corresponding to a \emph{virtual} and a \emph{real} bonding-curve phase.
In both cases, prices are determined by a deterministic mapping between the reserves of the token and Solana, defined by the constant-product invariant governing market dynamics.

The first environment is a \textit{virtual AMM}, which governs the early phase of each token lifecycle and implements the launchpad mechanism.
During this phase, trading takes place along a virtual bonding curve, described in detail below, which determines token prices algorithmically as a function of the cumulative amount of Solana committed by traders.
The bonding curve is initialized with protocol-defined virtual liquidity parameters and starts with zero real SOL reserves.
Once the total amount of Solana accumulated along the curve reaches a fixed threshold—known as the \textit{graduation point}—the token exits the launchpad phase and migrates to the second environment. 

The post-graduation phase takes place on a constant-product AMM with real reserves, PumpSwap.
Upon graduation, the protocol creates a canonical on-chain liquidity pool in which the graduated token acts as the base asset and the quote asset is Wrapped SOL (WSOL).
The migration procedure transfers the real SOL accumulated during the launchpad phase to the quote-token account of the PumpSwap pool, where it is wrapped into WSOL.
Importantly, only \emph{real} reserves are migrated: the \emph{virtual} reserves used during the bonding-curve phase are synthetic state variables employed solely for pricing and are not transferred on-chain.
As a result, the WSOL liquidity that seeds the PumpSwap pool is generated endogenously by traders’ purchases rather than by an upfront SOL deposit from the token creator.

This transition effectively converts a speculative, launchpad-based virtual market into a standard decentralized exchange environment based on a constant-product AMM with only real existing tokens.
After graduation, trading follows the same economic and microstructural principles as established DEXs such as Uniswap and Raydium, where liquidity provision, arbitrage, and price discovery emerge from interactions among heterogeneous market participants~\cite{adams2021uniswap, Angeris2020CFMM, Zhang2020FormalCFMM, CapponiJia2022, Milionis2022}.

The following subsections describe in more detail the mechanisms underlying these two stages:
first, the \textit{Virtual AMM} implemented by the Pump.fun launchpad, and second, the \textit{Real AMM} called PumpSwap, where trading continues after graduation.

\subsection{Virtual AMM: Pump.fun Launchpad}
The operational mechanism of the Pump.fun launchpad unfolds in the following stages. First, an initial supply of \(  y^{\rm mint}_0 = 10^9 \) tokens is minted. This total supply is then divided into two parts: 
\( y^{\rm virt, mint}_0 = 0.7931 \times 10^{9} \) tokens are allocated to the \emph{virtual bonding curve}, while the remaining fraction \( y^{\rm pswap, mint}_0 = 0.2069 \times 10^{9} \) is 
reserved for a potential \emph{graduation} into a \textit{Real PumpSwap} AMM bonding curve. Note that the creator does not incur in any cost for the new token.

%only cost incurred by the token creator is the gas fee of the first swap, which is approximately equal to $2\$$, if the creator does it.

Subsequently, a \emph{virtual AMM} is created, so named because its initial liquidity consists of both 
virtual synthetic and real tokens. Specifically, the bonding curve is initialized with 
\(x^{\rm synt}_0 = 30\) virtual synthetic tokens SOL, \(y^{\rm virt, mint}_0 = 0.7931 \times 10^{9}\) real tokens, and 
\(y^{\rm synt}_0 = 0.2799 \times 10^{9}\) virtual synthetic tokens.
Hence, the total initial liquidity state can be expressed as
\[
x^{\text{tot,virt}}_0 = x^{\rm synt}_0 = 30 \ \text{SOL}, \qquad 
y^{\text{tot,virt}}_0 =  y^{\rm virt, mint}_0 + y^{\rm synt}_0 = 1.073 \times 10^{9} \ \text{tokens}.
\]

As other decentralized exchanges, this AMM works with a constant-product rule i.e. indicating the pool reserves at a generic time as \((x,y)\) it is always satisfied
\[
x\,y = k = x_0^{\rm tot,virt} y_0^{\rm tot,virt},
\]
so that the bonding curve is simply the set of points \((x,y)\) with the same invariant \(k\) and  the marginal price is $P=x/y$. However, differently from other AMMs, during the virtual bonding curve phase liquidity cannot be added or removed with mint and burn actions and it is therefore always equal to $k$.

\iffalse
This configuration is deliberately designed to prevent the price $P$ from diverging under the simple  constant-product rule,
\[
x \, y = k = x_0^{\rm tot,virt} y_0^{\rm tot,virt} \,\,,
\]
\[
P = \frac{x^{\rm tot,virt}}{y^{\rm tot,virt}} \,\,,
\]
which governs this virtual AMM.

Under the constant-product rule that governs the bonding curve,
the pool reserves \((x,y)\) always satisfy
\[
x\,y = k,
\]
so that the bonding curve is simply the set of points \((x,y)\) with the same invariant \(k\).
It is often convenient to define the pool liquidity as
\[
L = \sqrt{k} = \sqrt{x\,y},
\]
which is proportional to the geometric mean of the reserves.
 \fi

The transaction fee is applied externally to the pool and is charged on each buy or sell swap. 
Specifically, a fee equal to \( 1 - r = f = 1.25\% \) of the amount of exchanged  SOL is collected. 
In particular, the \( 0.3\% \) of the total amount is distributed to the token creator, while the remaining portion $0.95\%$ is  collected by the \texttt{Pump.fun} protocol as a platform fee.
When a trader sends an amount \(\Delta x\) of real SOL to the pool, only a fraction \(r \Delta x\) effectively enters the bonding curve, after fees.
Given the invariant \(k = x^{\rm tot,virt}_0 y^{\rm tot,virt}_0\), the new total SOL reserve is
\[
x^{\rm tot,virt} = x^{\rm tot,virt}_0 + r \Delta x,
\]
and the corresponding token reserve on the bonding curve is
\[
y^{\rm tot,virt} = \frac{k}{x^{\rm tot,virt}}.
\]
Hence, the number of tokens purchased and thus withdrawn from the pool is
\[
 \Delta y(\Delta x) 
  = y^{\rm tot,virt}_0 - y^{\rm tot,virt}
  = y^{\rm tot,virt}_0 - \frac{k}{x^{\rm tot,virt}_0 + r \Delta x}.
\]

This relation defines the cumulative quantity of tokens issued as a function of the Solana deposited.

By definition, graduation occurs when the total reserve \( x^{\mathrm{tot,virt}} \) reaches \( 115 \)~SOL. 
At that point, the transition to the real \textit{PumpSwap} AMM takes place by creating a new, real bonding curve using the remaining 
token supply \( y^{\mathrm{pswap,mint}}_0 \) and the portion of real SOL $x_0^{\rm pswap}$ from the virtual bonding curve, equal to \( 85 \)~SOL. 
At the graduation point, the price remains continuous by construction, i.e.:
\iffalse
\[
\lim_{x^{\mathrm{tot,virt}} \to 115} P(x^{\mathrm{tot,virt}}) 
= 
\lim_{x^{\mathrm{tot,pswap}} \to 85} P(x^{\mathrm{tot,pswap}}) \,\,,
\]
in fact:
\fi
\begin{align}
    \frac{115\,{\rm SOL}}{y^{\rm synt}_0} \simeq \frac{85\,{\rm SOL}}{y^{\rm pswap,mint}_0} \,\,,
\end{align}
Up to minor transaction costs, the migration process also consumes a small portion of the real SOL reserves to cover 
the \emph{gas fee} required for the creation of the new AMM pool. This fee is deducted directly from the real SOL 
balance accumulated within the virtual bonding curve.
Even though, the effective liquidity \(\sqrt{x \, y} \) undergoes a discrete drop due to the 
transition from virtual to real reserves.

\subsection{Real AMM: PumpSwap}

Upon graduation, the token transitions to the \textit{real AMM}, known as \textit{PumpSwap}.  
At this point, the system reclaims the 30~SOL virtual advance used to initialize the curve, effectively reducing the total SOL reserve from 115~SOL to 85~SOL.  
Simultaneously, the token reserve decreases from its virtual synthetic total $y^{\rm synt}$ to $y^{\rm pswap,mint}$ tokens, corresponding to the amount that would remain in the pool according to the bonding curve mechanics.  
Importantly, the token price at the moment of transition, $p_{\text{grad}}$, remains constant by design to ensure a smooth migration between the virtual and real environments.

In the real phase, a liquidity pool is created in the PumpSwap AMM with the pair (85~SOL, $2.069 \times 10^8$ tokens), defining the initial market price:
\[
p_{\text{grad}} = \frac{85}{2.069 \times 10^8}.
\]
From this point onward, the token becomes fully tradable on-chain, and its price evolves freely according to the dynamics of the real \textit{PumpSwap} AMM.  
This second phase resembles a standard decentralized exchange, where liquidity providers and traders interact continuously, and price movements are determined by actual supply and demand.

The two-phase architecture of Pump.fun thus serves a dual purpose: the virtual AMM efficiently bootstraps liquidity and price discovery in a deterministic way, while the real \textit{PumpSwap} AMM enables open trading under true market conditions.  
This structure explains why Pump.fun can sustain a very high volume of token launches and rapid trading activity, a feature made possible by Solana low latency and high transaction throughput.

\section{The Dataset}

The investigated dataset covers a one-month period, from September 1 to October 1, 2025, and includes only newly created tokens. During this time, a total of 655,770 tokens were created by 243,123 distinct token creator addresses. Among these, only 4,338 coins reached the graduation point, corresponding to a graduation frequency of approximately 0.63\%. In total, 2,600,790 distinct trader addresses were recorded, although the effective number of unique users interacting with the platform is likely lower due to the use of multiple wallets per trader. %An informative indicator is the ratio between the number of distinct trader addresses and the total number of trades per coin, which provides a normalized measure of market participation {\bf FL: E dunque? Lo studieremo piu' avanti?}.

Data were collected directly from the Solana blockchain by parsing on-chain data from the Pump.fun and PumpSwap programs. This allowed us to track all newly created tokens across both the virtual and the real bonding curves. We retrieved and decoded all transaction events associated with token creation, bonding-curve interactions, and swap operations. On this launchpad system, i.e. on the virtual bonding-curve, adding and removing liquidity is not allowed by the protocol. Each transaction was mapped to its corresponding token launch, enabling the reconstruction of time-resolved bonding-curve states and the cumulative amount of SOL committed to each project. This approach ensures full transparency and reproducibility of the dataset, while accurately capturing the early trading and funding dynamics of the launchpad.

Each record in the dataset corresponds to a single on-chain transaction and includes the following core variables:

\begin{itemize}
    \item \texttt{timestamp} and \texttt{local\_time}: blockchain and local time of the transaction, used for temporal alignment and intra-block sequencing;
    \item \texttt{signature}: unique transaction hash identifying each operation;
    \item \texttt{mint}: token address associated with the launched token;
    \item \texttt{coin\_creator}: address of the token creator;
    \item \texttt{trader}: wallet address of the trader executing the transaction;
    \item \texttt{txType}: categorical variable describing the transaction type (\textit{create}, \textit{buy}, \textit{sell});
    \item \texttt{inBondingCurve}: boolean flag indicating whether the transaction occurred during the bonding curve phase;
    \item \texttt{vSolInBondingCurve} and \texttt{vTokInBondingCurve}: current amounts of SOL and tokens locked in the virtual bonding curve just after the transaction;
    \item \texttt{solAmount} and \texttt{tokenAmount}: quantities of SOL and tokens involved in the specific trade;
    \item \texttt{isBot}: binary indicator (0/1) identifying bot-like activity, derived from log-level inspection of transaction sources.

\end{itemize}

Regarding the variable isBot, we extracted this information by looking at the logs of the transactions on the blockchain. Specifically, each transaction recorded on the Solana blockchain contains metadata about the calling program and the origin of the request.  
By analyzing these logs, we distinguished between two cases: (i) transactions routed through the official Pump.fun graphical interface (a website), and (ii) transactions directly invoking the smart contract without passing through the platform’s frontend.  
The former are typically associated with automated or scripted interactions, while the latter correspond to manual trades executed via the public user interface.  
Accordingly, we assign \textit{isBot} = 1 to transactions that directly interact with the contract (bypassing the interface) and \textit{isBot} = 0 to those executed through dApp interfaces.  
This classification allows us to isolate algorithmic or non-standard trading behaviors and assess their impact on the probability of graduation.

This structure provides both micro-level transactional granularity and macro-level aggregation possibilities, enabling the estimation of conditional graduation probabilities as a function of bonding curve dynamics and trader composition.

\section{Descriptive statistics of the investigated dataset}
\begin{figure}
\centering
\includegraphics[width=1\textwidth]{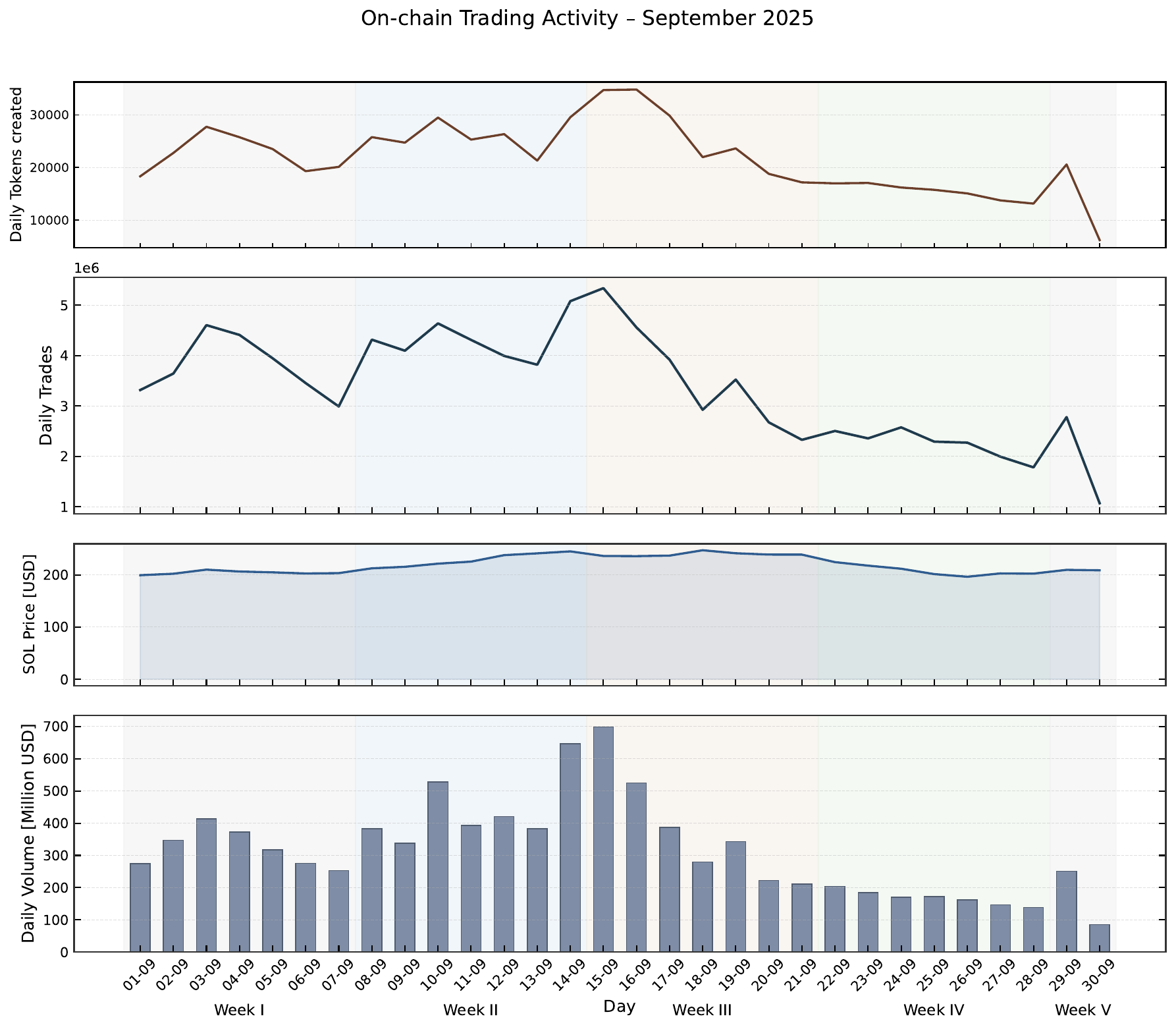}
\caption{On-chain trading activity in the PumpFun ecosystem during September 2025.
The top panel reports the daily number of newly created tokens.
The second panel shows the daily number of trades.
The third panel displays the average daily price of Solana (SOL) in USD.
The bottom panel reports the daily trading volume, expressed in million USD, computed from SOL-denominated volumes rescaled by the corresponding daily average SOL price.}
\label{fig:market}
\end{figure}

\begin{figure}[H]
\centering
\includegraphics[width=0.6\textwidth]{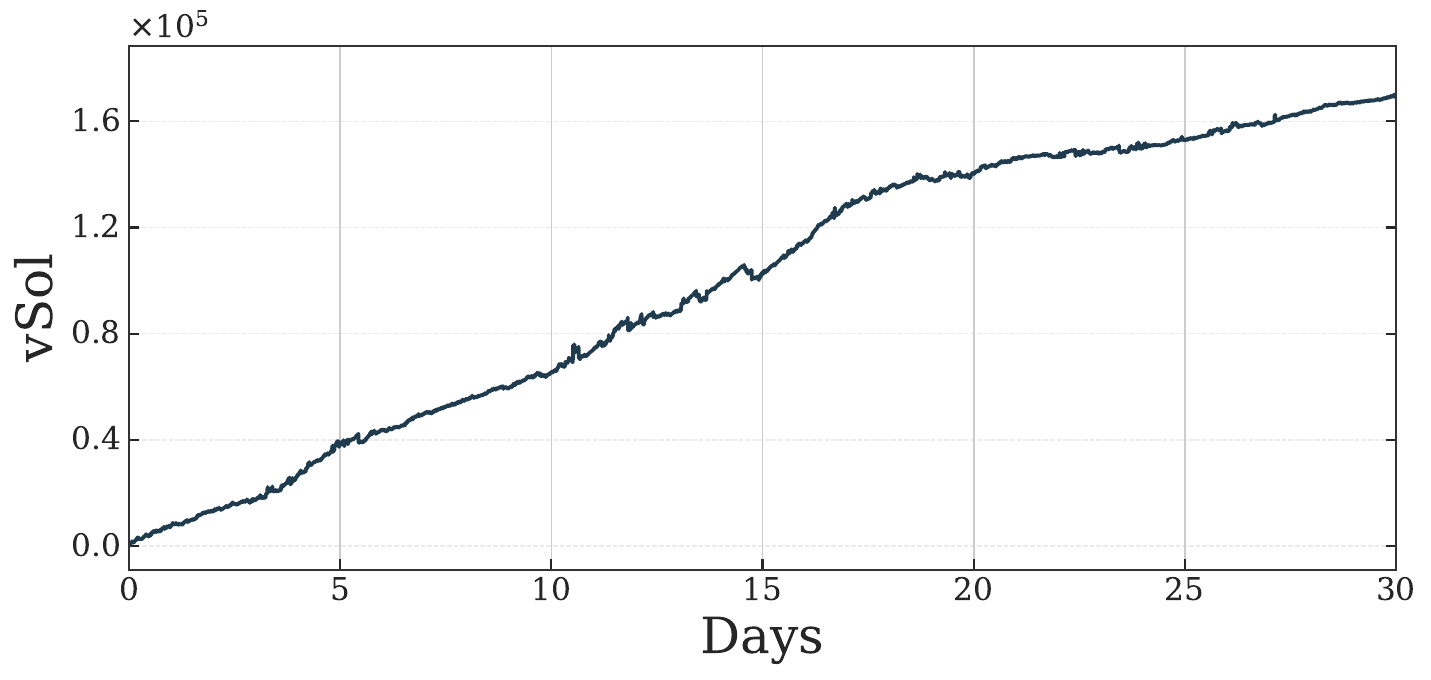}
\caption{Cumulative signed amount value of Solana (vSol) circulating through all newly created tokens during September 2025, including both bonding curve and post-graduation AMM activity.}
\label{fig:vsol_time}
\end{figure}

This section provides an overview of the main aggregate properties of the market captured in our dataset, in order to illustrate its scale and overall activity.
Figure~\ref{fig:market} provides an initial macroscopic view of the ecosystem during the observation month, summarising four dimensions of activity. From the top:  daily tokens created, daily number of trades, average USD price of Solana, and daily USD-denominated trading volume. 
The daily number of trades shows pronounced fluctuations, with activity ranging from roughly 2 to 5 million trades per day. From the figures, we observe an obvious positive correlations between the trade activity and the number of new tokens created.
The third panel indicates that the price of Solana remained comparatively stable throughout the period. The relative smoothness of the price trajectory suggests that variations in market activity are not primarily driven by exogenous price shocks.
The bottom panel shows that the daily  volume dynamics broadly mirror those of the trade count, with particularly intense activity during the central part of the month, where several days exceed 600--700 million USD in exchanged value. After mid-month, the USD volume progressively decreased, stabilizing at significantly lower levels in the last week. The combination of somewhat stable SOL prices and sharply varying USD volume underscores that fluctuations in the Pump.fun ecosystem activity are predominantly driven by changes in trading intensity rather than price movements of Solana.
As explained above, a graduated token must reach a threshold of approximately $85\,\mathrm{SOL}$ on its bonding curve. Considering the $4{,}338$ tokens that successfully graduated, this corresponds to a total net exchanged volume of
\[
85\,\mathrm{SOL} \times 4{,}338 \simeq 3.7 \times 10^{5}\,\mathrm{SOL}
\simeq 7.4 \times 10^{7}\,\mathrm{USD},
\]
where we consider only graduated tokens. To provide an aggregate, directly observable measure of the economic relevance of this ecosystem, we construct a market-level flow metric. Specifically, in Fig.~\ref{fig:vsol_time} we compute the total net SOL cash flow generated by all newly created tokens over the month, accounting both for the bonding-curve phase on Pump.fun and for the post-graduation trading on AMMs. This quantity captures how much SOL is, on net, absorbed and retained by the Pump.fun token market over the sample period, and thus offers a simple proxy for the scale and importance of this segment of on-chain activity.

At the end of September, this cumulative value reached approximately $1.6\times10^5$~SOL $\simeq 3.2 \times 10^7$ \$, which can be interpreted as the residual amount of Solana effectively created and remaining within the system at the end of the observation period.
This measure therefore captures the aggregate liquidity that continues to exist within the system, integrating both pre- and post-graduation dynamics. We also note that this aggregate net flow is of the same order of magnitude as the SOL that can be transferred into canonical AMM liquidity at graduation. This pattern may be related to post-graduation persistence: a subset of graduated tokens remains actively traded over time and, since liquidity-pool creation on Solana AMMs is permissionless, additional pools may be created on other DEX venues beyond the canonical post-graduation market.  In the most extreme cases, sufficiently popular tokens are also listed on centralized exchanges (e.g., from Pump.Fun all data history: POPCAT on Binance.US; GOAT via Binance futures; PNUT on major CEXs such as Gate and OKX).

\begin{figure}[H]
\centering
\includegraphics[width=0.6\textwidth]{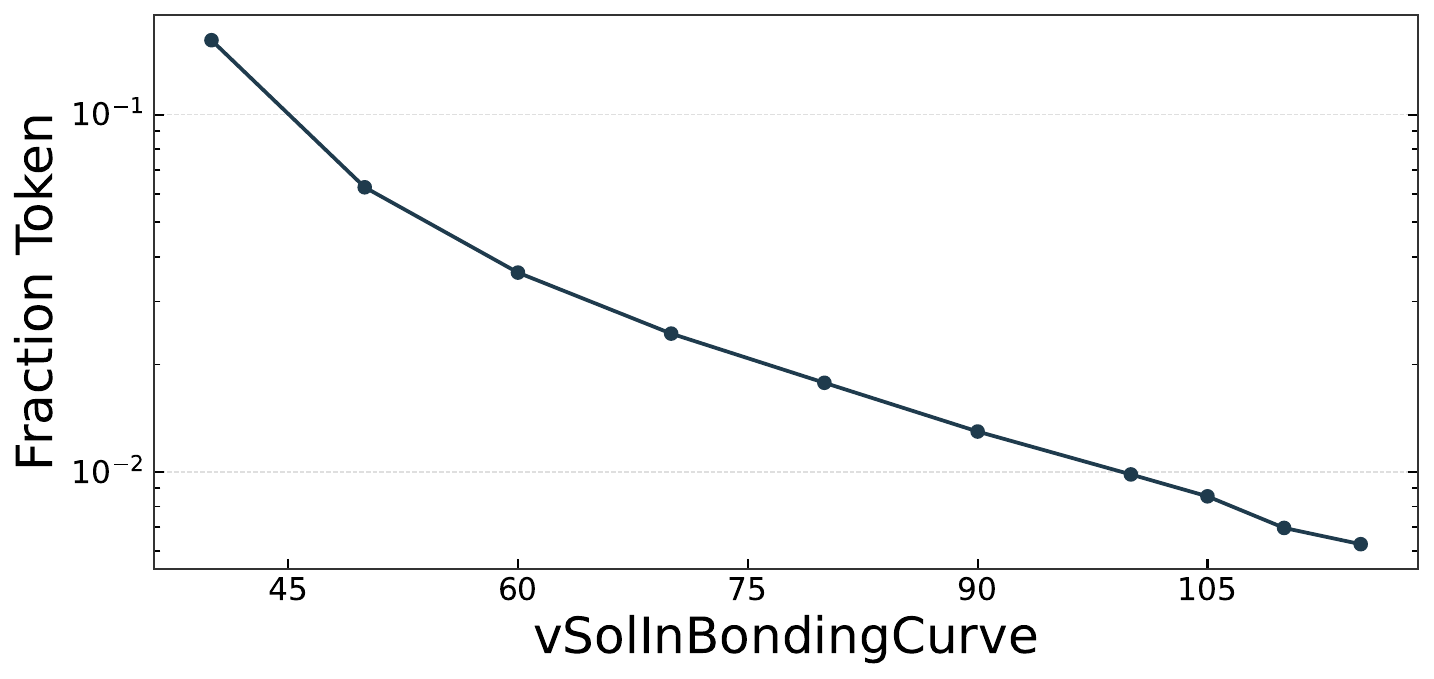}
\caption{Empirical survival curve of bonding-curve progress.
For each threshold of SOL in bonding curve, we report the number of newly created tokens reaching that amount.
The ordinate is shown on a logarithmic scale, highlighting the rapid drop-off in the fraction of tokens that accumulate large amounts of SOL along the bonding curve.}
\label{fig:token_number}
\end{figure}

Beyond aggregate liquidity, it is also interesting to quantify how stalling tokens progress along the bonding curve before graduation (see Figure~\ref{fig:token_number}) 
%provides a compact view of how far tokens typically progress along the bonding curve before stalling. 
For each level of \textit{vSol} in the bonding curve, we count how many tokens ever reach at least that value during their lifetime. 
The resulting curve is monotonically decreasing by construction and displays a steep decay: while a large mass of tokens reaches only low \textit{vSol} levels, only a tiny fraction approaches the upper part of the curve near the graduation region. 
On the logarithmic scale, the approximately linear behavior over an intermediate range suggests a fast, nearly exponential drop in the probability of reaching higher capital thresholds, consistent with a strong selection effect in early trading and funding dynamics.

\begin{figure}
\centering
\includegraphics[width=0.9\textwidth]{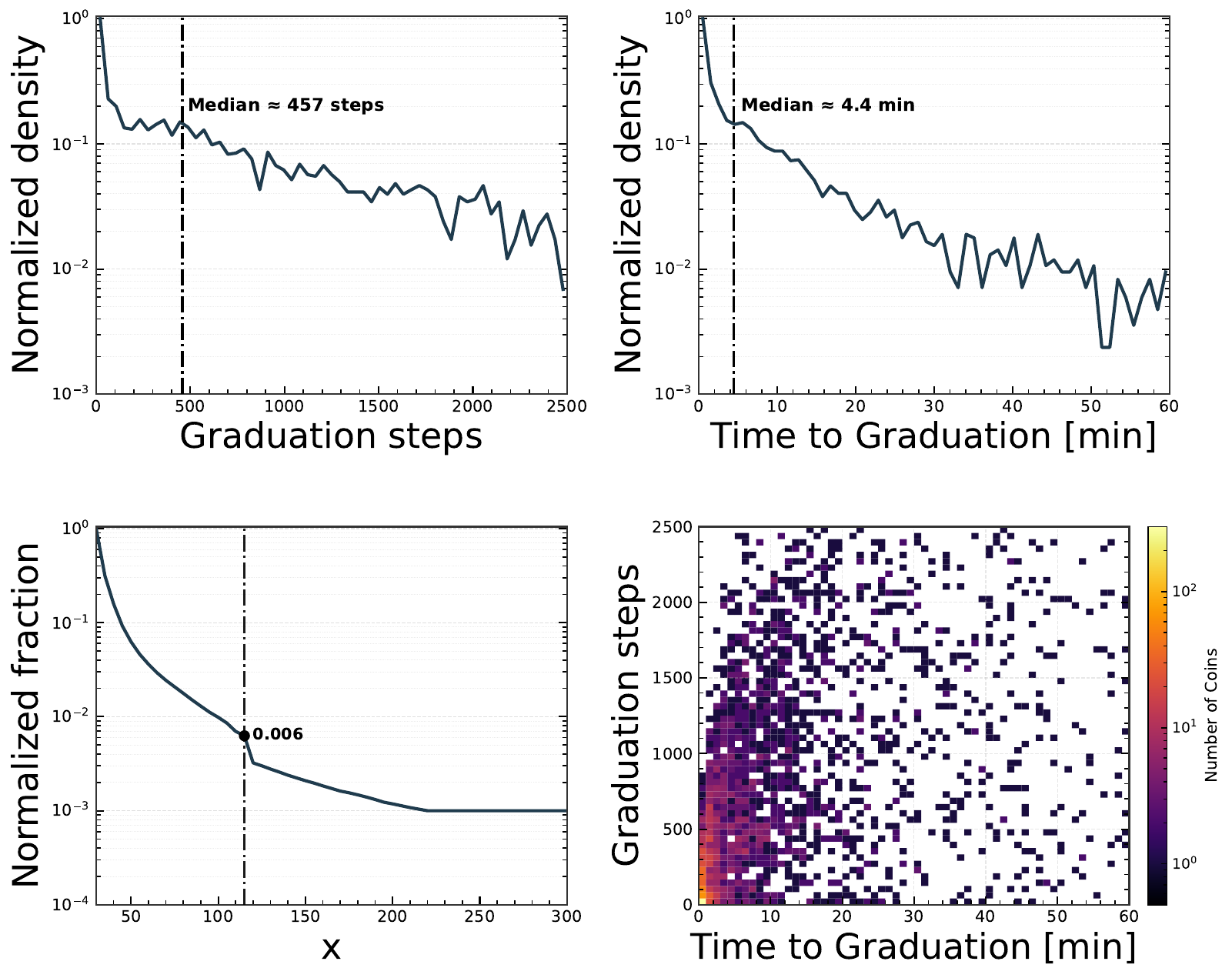}
\caption{
Graduation dynamics of Pump.fun tokens across multiple dimensions.
\textbf{Top left}: Empirical distribution of the number of bonding-curve steps required to reach graduation.
\textbf{Top right}: Empirical distribution of time to graduation (in minutes) from token creation.
Both distributions are strongly right-skewed, with medians of approximately $457$ steps and $4.4$ minutes, respectively (dashed lines).
\textbf{Bottom left}: Normalized fraction of tokens whose maximum bonding-curve state exceeds a given value of \textit{vSol}, showing that only a small fraction of launched tokens reaches the graduation region and migrates to the real \textit{PumpSwap} AMM pool.
\textbf{Bottom right}: Joint distribution of graduation steps and time to graduation, with color intensity indicating the number of tokens on a logarithmic scale.
The concentration of mass at short times and low step counts highlights the fast and momentum-driven nature of successful launches, while the sparse tail reflects heterogeneous and less efficient paths to graduation.
}
\label{fig:vsol_threshold}
\end{figure}

The two upper panels in Figure~\ref{fig:vsol_threshold} report information on the steps and the time tokens that are usually required to graduate. The top left panel reports the empirical distribution of graduation steps, i.e.\ the number of on-chain bonding-curve updates (swaps/events) required for a token to reach the graduation threshold. In fact, on the Pump.fun launchpad, the only event , which drives measurable market changes, is the swap event.
The distribution is strongly right-skewed: most tokens graduate after a relatively small number of steps, while a long tail extends to several thousand steps. 
The vertical dashed line marks the median, $\simeq 457$ steps, indicating that half of the graduated tokens reach migration within a few hundred bonding-curve events.

The top right panel shows the corresponding distribution of time-to-graduation, measured in minutes from token creation to the graduation transaction. 
Also in this case the distribution exhibits a pronounced heavy tail: the bulk of graduations occurs very rapidly, whereas a minority of tokens requires tens of minutes to migrate. 
The median time to graduation is $\simeq 4.4$ minutes (dashed line), highlighting the high-frequency and strongly momentum-driven nature of successful launches on Pump.fun.

The bottom-left panel shows that only a small fraction of the total number of tokens actually reaches the graduation threshold and migrates to the real \textit{PumpSwap} AMM pool.
Specifically, the curve reports the normalized fraction of tokens whose bonding-curve trajectory exceeds a given value of \textit{vSol}.
The sharp decay highlights that the vast majority of tokens stalls at relatively low levels of accumulated Solana, while only about $0.6\%$ of all launched tokens manage to graduate.

The bottom-right panel provides a joint view of the graduation process by displaying the empirical density of tokens in the plane defined by time to graduation and number of graduation steps.
Color intensity encodes the number of tokens in each bin on a logarithmic scale.
The plot reveals a strong concentration of mass at short times and low step counts, indicating that most successful tokens graduate very quickly and with relatively few bonding-curve interactions.
At the same time, a broad and sparse tail extends toward longer times and larger numbers of steps, reflecting heterogeneous and less efficient paths to graduation for a minority of tokens.
%{\bf MI PARE CHE LA CURVA IN BASSO A SINISTRA DICA LA STESSA COSA DI FIGURA 3. INOLTRE IL PANNELLO IN BASSO A DESTRA NON MI PARE MOLTO INFORMATIVO (SI PUO' DIRE A PAROLE). FORSE POTREMMO LASCIARE SOLO I TOP PANELS DI FIGURA 4.}

\section{Conditional graduation probability}

\begin{figure}
\centering
\includegraphics[width=0.6\textwidth]{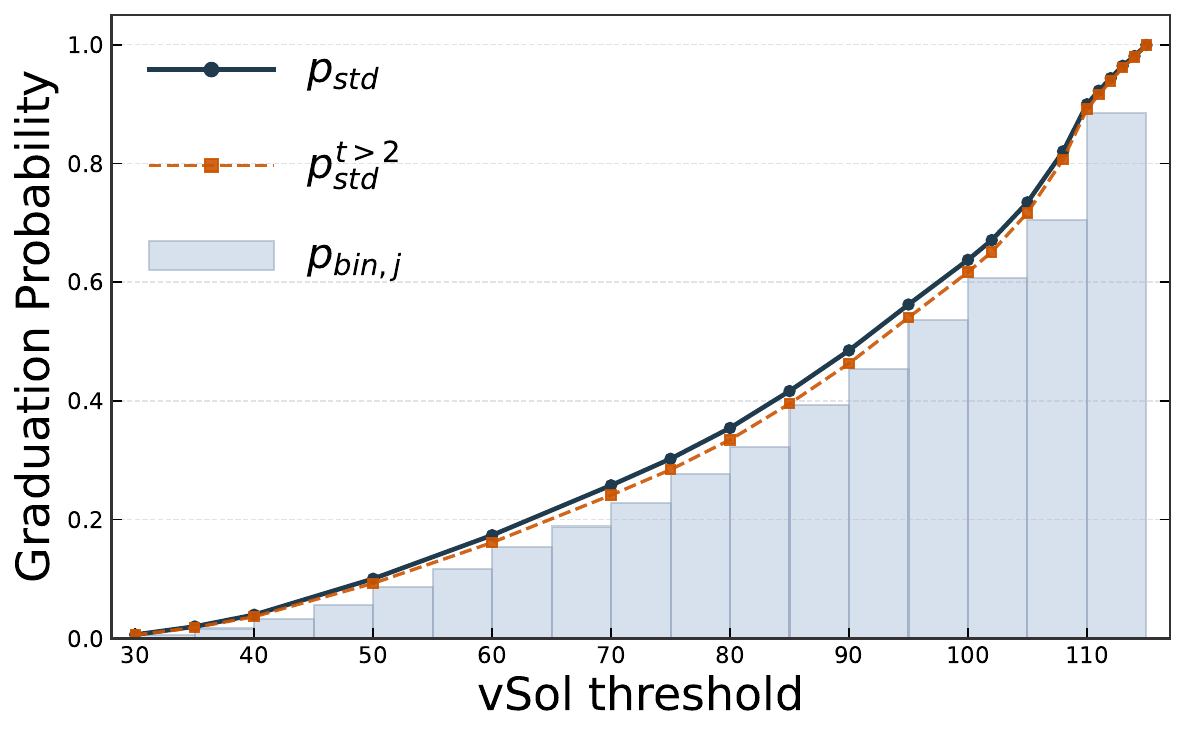}
\caption{Empirical probability of graduation as a function of the virtual Solana threshold (\textit{vSol}).
The solid line ($p_{\mathrm{std}}$) reports the baseline estimator $P(\mathrm{grad}\mid \exists\,x:x>vSol)$.
The dashed line ($p_{\mathrm{std}}^{t>2}$) shows the same probability after excluding ultra-fast graduations with time to graduation shorter than two seconds.
Light-blue bars ($p_{\mathrm{bin},j}$) display a non-parametric binned estimate over intervals of width $5$~SOL.
The close agreement between the curves indicates that extremely rapid, likely bot-driven events do not materially affect the graduation-probability profile, while the monotonic increase reflects the deterministic progression of tokens along the bonding curve toward the graduation threshold. 
}
\label{fig:gradvsol}
\end{figure}

In the following, we exploit this structure to study how the probability of graduation evolves as a function of information that is observable in real time during a token’s launch phase.

As a first step, we focus on the most natural state variable of the launch mechanism itself: the amount of Solana currently locked in the bonding curve.
We therefore begin by estimating the probability of graduation conditional on the current level of virtual Solana (\textit{vSol}), before moving on to richer conditioning sets that capture additional dimensions of trading behavior and agent composition.

We estimate this probability empirically by computing, for each level of \textit{vSol} observed across all tokens, the fraction of coins that ultimately graduate.
More precisely, for any threshold on the bonding curve, we consider the set of tokens that have reached at least \textit{vSol} at some point in their life cycle and measure the proportion of those that eventually graduate.
The resulting conditional probabilities are reported in Fig.~\ref{fig:gradvsol}.

The solid blue curve represents our baseline estimator, 
\[
p_{\mathrm{std}}(vSol) = P\bigl(\mathrm{grad} \mid \exists\, x : x > vSol \bigr),
\]
namely the probability that a token eventually graduates conditional on having crossed a given \textit{vSol} level.
This quantity provides a smooth, global characterization of how the likelihood of graduation increases as more Solana is accumulated along the bonding curve, and it will serve as the reference object throughout the remainder of the paper.

The dashed orange curve reports the same conditional probability after excluding extremely fast graduations, defined as events with a total time to graduation shorter than two seconds:
\[
p_{\mathrm{std}}^{t>2}(vSol) = P\bigl(\mathrm{grad} \mid \exists\, x : x > vSol,\; t_{\mathrm{grad}} > 2~\mathrm{s}\bigr).
\]
This filter removes ultra-fast events that are likely driven by automated bots or anomalous bursts of activity.
The close alignment between $p_{\mathrm{std}}^{t>2}$ and $p_{\mathrm{std}}$ indicates that such events do not materially affect the shape of the graduation-probability curve, supporting the robustness of the baseline estimator.

For completeness, the light-blue bars report a non-parametric binned estimate of the same probability.
For each interval $[vSol,vSol+5)$ of width 5~SOL, we compute the fraction of tokens that graduate among those whose bonding-curve trajectory enters that bin.
This discretized representation provides a direct view of the underlying empirical frequencies and visually encodes the local statistical support, complementing the smoother baseline curve.

As expected, all estimators display a monotonically increasing probability of graduation as \textit{vSol} grows, with values approaching one near $115$~SOL.
This level corresponds to the effective graduation threshold, consisting of $30$ virtual SOL (initialized at creation) and $85$ real SOL deposited by traders.
Once the bonding curve accumulates this total amount, graduation becomes virtually guaranteed and the empirical probabilities converge to unity.

In the remainder of the paper, we use $p_{\mathrm{std}}(vSol)$ as the breakeven graduation-probability curve and study how it shifts when conditioning on additional variables capturing trading intensity, agent composition, and behavioral heterogeneity.

However, the predictive information contained in \textit{vSol} alone is limited: tokens with similar \textit{vSol} values may exhibit very different outcomes depending on behavioral and structural factors that are not captured by aggregate capital accumulation.
This motivates the introduction of additional conditioning variables, aimed at disentangling how trading intensity, agent composition, and participation patterns affect the likelihood of graduation.

Before turning to these richer conditioning sets, it is useful to place the graduation-probability curves in an explicitly economic perspective.
In particular, one may ask whether the empirical probabilities we estimate are large enough to be economically meaningful for a trader facing the mechanical price dynamics implied by the bonding-curve design.
To this end, we first introduce a simple breakeven that maps graduation probabilities into a minimal profitability condition under a naive buy-and-hold strategy.
This breakeven provides a transparent reference scale against which the magnitude and relevance of the estimated probabilities can be assessed.

%Having established this economic baseline, we then study how conditioning on additional variables shifts the graduation-probability curves relative to the breakeven, thereby clarifying which features of early trading activity are not only statistically predictive, but also economically significant.

\subsection{An economic breakeven curve}

The conditional graduation-probability curves introduced above naturally raise the question of whether such information can be exploited in an economically meaningful way.
Rather than proposing a realistic or optimized trading strategy, we use a deliberately simple breakeven to translate graduation probabilities into a minimal profitability condition.
This exercise serves as a sanity check: it allows us to assess whether the empirical likelihood of graduation is, in principle, strong enough to compensate for the mechanical price dynamics implied by the bonding-curve design.

Specifically, imagine that by considering additional features $\theta$, we estimate the graduation probability $p(vSol;\theta)$ which depends also on the current $vSol$ and hence on the price $vSol^2/k$. We consider a simple buy-and-hold trading strategy where we buy a given token when $vSol$ first reaches a chosen threshold (corresponding to the candidate $x$-axis entry points in the figures), and we sell only at graduation, i.e., when $vSol = 115$ and the price is $115^2/k$.
If the token never reaches graduation, we never sell (the position remains open) and we lose our investment\footnote{
Throughout, we ignore gas costs as well as protocol and creator fees, which are negligible under the small-fee approximation $f = 0.0125 \ll 1$. }. %Suppose we invest 1 SOL when the bonding curve contains an amount $vSol$ of SOL. Then, with probability $p(vSol)$ graduation occurs and the investment yields a positive return equal to the SOL price difference between the graduation price and 1 SOL. With probability $1 - p(vSol)$ the investment results in a total loss. Thus, the expected PnL is
The expected simple return $r$ of the strategy is
\begin{align}
    & {\mathbb E}[r|vSol;\theta] = %E[{\rm r}]  = 
    \Biggl(\frac{115^2}{k}\frac{k}{vSol^2}-1\Biggl)  p(vSol; \theta) + (0 - 1)  \Bigl( 1 - p(vSol;\theta) \Bigl) \,\,.
\end{align}
The strategy is economically profitable if ${\mathbb E}[r|vSol;\theta]>0$ hence if
\begin{align}    
 p(vSol;\theta)  > \frac{vSol^2}{115^2} \,\,. 
\end{align}

Thus, if the conditional graduation probability curve $p(vSol;\theta)$  lies above the parabolic boundary ${vSol^2}/{115^2}$, the naive buy-and-hold strategy based on it delivers a positive expected profit (under the same cost assumptions). We call the curve $p={vSol^2}/{115^2}$ as the {\it breakeven curve}. It is important to stress that this is the breakeven curve only for buy-and-hold strategies. In other words, it might be possible to devise dynamic strategies delivering positive expected profits also when the conditional probability lies below the breakeven curve. When there are no additional features $p(vSol;\theta)=p_{std}(vSol)$. As shown below (see for example Fig. \ref{fig:pgrad_isbot}) $p_{std}(vSol)$ lies below the breakeven curve, hence it is not possible to make profits with a buy-and-hold strategy based only on $vSol$ (or the price).

In the next section, we will systematically compare the empirically estimated graduation-probability curves with this breakeven boundary in order to quantify how different conditioning variables $\theta$ shift the probability mass relative to the profitability threshold.

\section{Prediction variables}

We consider four main variables that capture distinct aspects of market behavior and participant profiles and might be useful in improving success prediction:

\begin{itemize}
    \item \textbf{Bot activity (\textit{isBot})}: a binary indicator identifying transactions associated with algorithmic trading accounts.  
    Conditioning on this variable reveals whether coins traded primarily by bots are more or less likely to graduate for a given \textit{vSol}.
    
    \item \textbf{Number of trades}: the cumulative number of transactions recorded for a token up to a given \textit{vSol}.  
    This variable proxies market activity and engagement, allowing us to evaluate whether a higher trading frequency increases the probability of graduation.
    
    \item \textbf{Presence of successful traders}: a categorical variable marking whether at least one wallet that achieved high profits during the month has traded the token.  
    Conditioning on this variable helps assessing whether the involvement of successful or possibly informed traders correlates with higher graduation chances.
    
    \item \textbf{Top token creators}: an indicator for whether the token was issued by one of the most prolific creators, defined as those who launched the largest number of tokens in the observation period.  
    This conditioning captures potential experience or network effects among frequent creators.
\end{itemize}

Each of these variables is used to compute a new conditional graduation probability curve, estimated in the same way as the baseline in Figure~\ref{fig:gradvsol}.  
In the following subsections, we compare these curves to the baseline case and discuss how the inclusion of each conditioning variable modifies the probability landscape, thereby improving the predictive understanding of graduation events.

\subsection{Conditioning on the share of bot-like activity}

A key dimension of heterogeneity in the Pump.fun market concerns the distinction between algorithmic and manual trading behavior.  
To capture this aspect, we define the variable \textit{isBot} as a binary indicator (0/1) identifying bot-like activity, derived from log-level inspection of transaction sources.  
Specifically, each transaction on the Solana blockchain contains metadata about the calling program and the origin of the request.  
By analyzing these logs, we distinguished between transactions routed through the official Pump.fun frontend (the website) and those directly invoking the smart contract on-chain.  
The latter interactions, which bypass the graphical interface, are typically associated with automated or scripted trading behavior.  
Accordingly, we assign \textit{isBot} = 1 to direct on-chain calls and \textit{isBot} = 0 to transactions executed via the Pump.fun website.

In this analysis, rather than conditioning on individual transactions, we aggregate information at the token level.
For each token, we compute the fraction of trades executed by addresses flagged as \textit{isBot} = 1 and use this quantity as a proxy for the presence of automated or sophisticated trading activity.
We then partition tokens into two groups depending on whether their bot share exceeds a given threshold, distinguishing markets with a high prevalence of bot-like activity from those dominated by non-automated, retail-like traders.

Formally, let $\tau$ denote the fraction of non bot-initiated trades for a given token, and let $\theta$ be a chosen threshold.
We define the conditional graduation probability 
\begin{equation}
    p_{\theta}(\mathrm{vSol}) = P\bigl(\mathrm{grad} \mid \tau > \theta\bigr),
\end{equation}
that is, the probability of graduation conditional on the token exhibiting a non bot share above $\theta$. In other words, the threshold $\theta$ represents a lower bound on the fraction of non sophisticated traders acting on the considered tokens.

Crucially, in order to preserve a causal and predictive interpretation, the non bot-share $\tau$ is computed using only trades that occurred up to the current level of \textit{vSol} on the bonding curve.

In other words, at each value of accumulated Solana, the conditioning variable is constructed exclusively from information that would have been observable to an external agent at that point in time.
This ensures that the resulting conditional probability curves capture genuine predictive content rather than ex-post correlations driven by future trading activity.

\begin{figure}[t]
\centering
\includegraphics[width=0.6\textwidth]{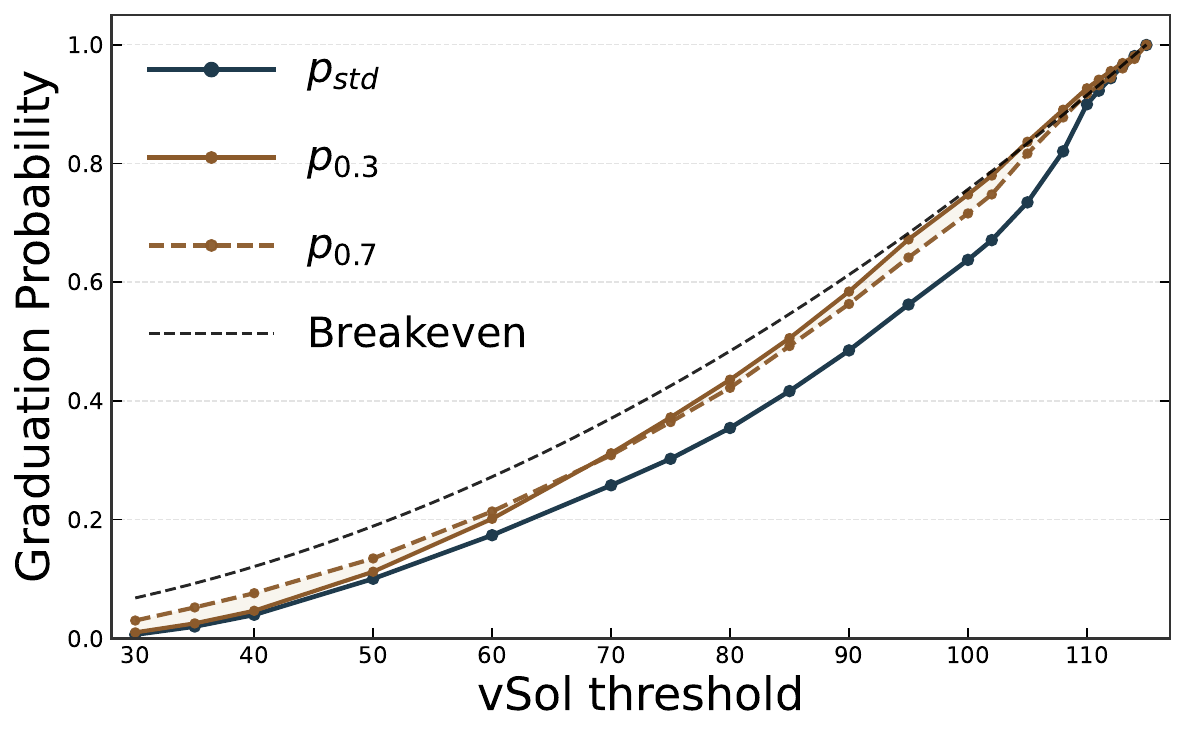}
\includegraphics[width=0.6\textwidth]{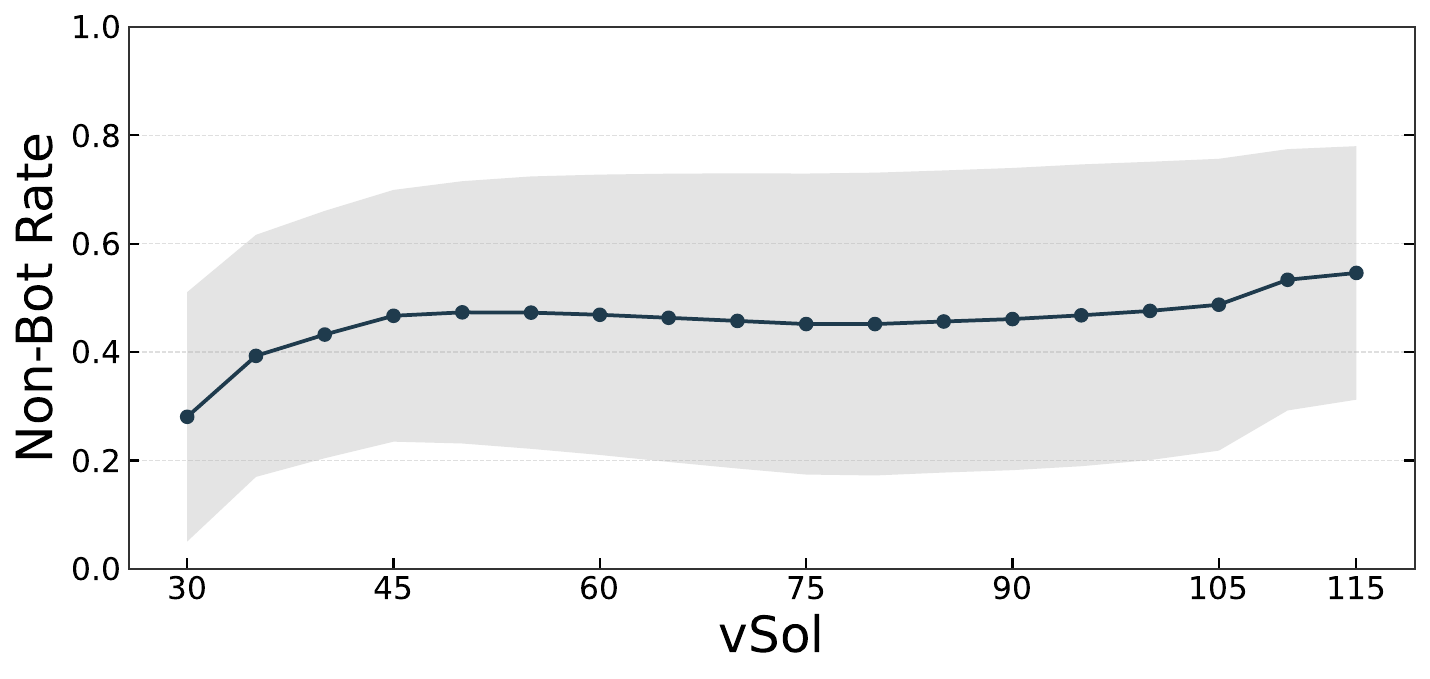}
\caption{The top panel shows the graduation probability as a function of the bonding-curve state variable \textit{vSol}, conditional on the intensity of non bot-like trading activity.
The solid blue line ($p_{\mathrm{std}}$) reports the baseline graduation probability.
The solid and dashed brown lines show the conditional probabilities for tokens with a share of non-bot-attributed trades above $0.3$ and $0.7$, respectively.
The black dashed line represents the breakeven profitability boundary derived from the naive entry-and-hold strategy.
Curves lying above the breakeven indicate regions in which the implied graduation probability is sufficient to offset the mechanical price increase along the bonding curve, yielding non-negative expected returns under the breakeven assumptions.
The bottom panel shows the mean bot rate as a a function of the price level. This serves as a motivation for choosing the extreme bot-probability values of 0.3 and 0.7.
}
\label{fig:pgrad_isbot}
\end{figure}

Figure~\ref{fig:pgrad_isbot} compares the baseline graduation-probability curve $p_{\mathrm{std}}$ with conditional curves obtained by partitioning tokens according to the share of bot-initiated trades.
Specifically, we consider two thresholds for the minimum presence of non-sophisticated traders, namely $\theta = 0.3$ and $\theta = 0.7$. We choose these extreme values because the average non-bot rate in our data lies between them; moving further away would not materially affect the implied probabilities, since only a small fraction of coins fall more than one standard deviation from the mean (see bottom plot in Fig. \ref{fig:pgrad_isbot}).
This means that we condition the graduation probability on tokens that, upon reaching a given SOL threshold on the bonding curve, exhibit at least $30\%$ of non-sophisticated traders (i.e.\ at most $70\%$ bot activity) and, respectively, at least $70\%$ of non-sophisticated traders (i.e.\ at most $30\%$ bot activity).
For reference, the dashed line reports the economic breakeven curve derived from the naive buy-and-hold strategy.

What emerges from the analysis is that both conditional curves lie systematically above the $p_{\rm std}$ curve. This indicates that incorporating information on the composition of traders participating in the price-formation process is highly informative.
In particular, the results suggest that a large participation of non-sophisticated traders is associated with an increased likelihood of reaching graduation.
A plausible interpretation is that sophisticated traders, including token creators, often engage in short-horizon strategies such as pump-and-dump schemes—which we analyze in detail in a later section—that tend to accelerate the depletion of liquidity and contribute to an early termination of the token lifecycle. We also stress that the two conditional graduation-probability curves, associated with the two different values of $\theta$, become statistically similar. Beyond that range, increasing the share of non-bot traders does not materially change the predicted probability: the signal carried by the non-bot percentage is essentially exhausted at high non-bot presence.
Finally, we observe that all conditional graduation-probability curves remain below the economic breakeven over most of the \textit{vSol} range, with the notable exception of the $p_{0.3}$ curve, which approaches the breakeven in the vicinity of the graduation threshold.
It should be emphasized that this comparison is purely qualitative.
The breakeven curve is derived from a deliberately naive buy-and-hold strategy and relies on strong simplifying assumptions, in particular the neglect of protocol, creator, and gas fees, which are assumed to be negligible.
As a result, the purpose of this exercise is not to assess realistic profitability, but rather to provide an interpretable reference scale against which the magnitude of the estimated graduation probabilities can be evaluated.

\subsection{Conditioning on trading intensity}

\begin{figure}
\centering
\begin{minipage}{0.48\textwidth}
    \centering
    \includegraphics[width=\textwidth]{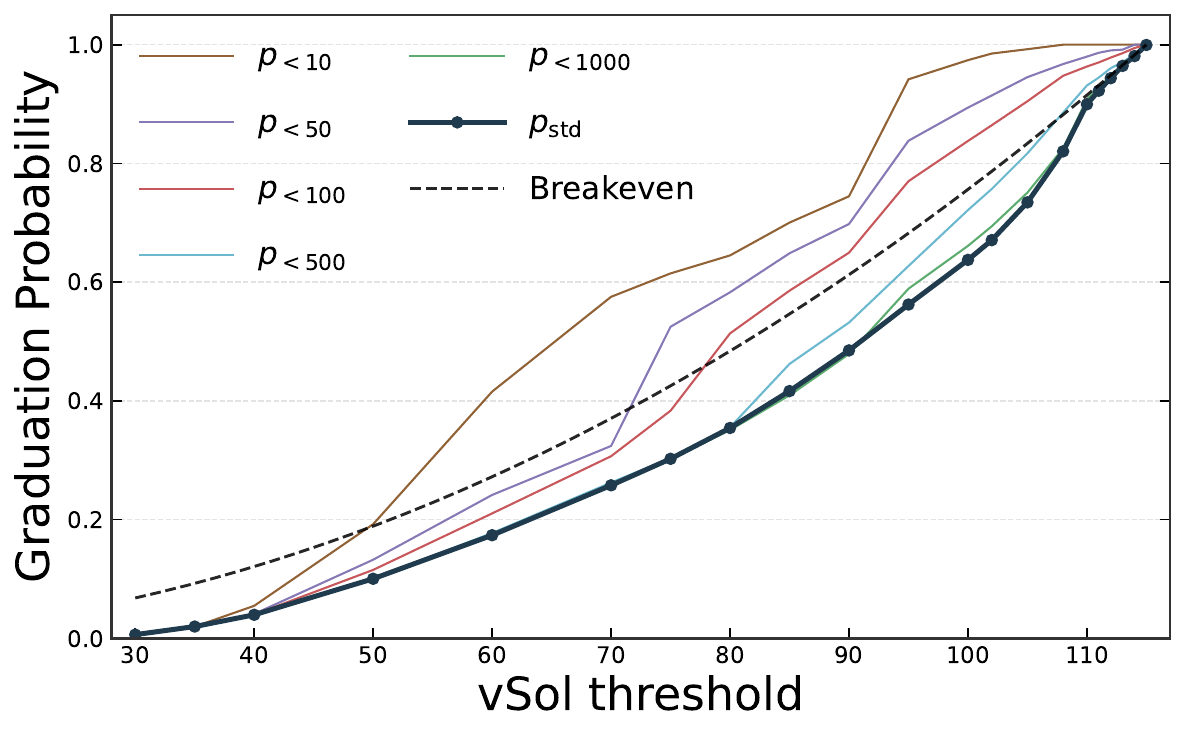}
\end{minipage}
\hfill
\begin{minipage}{0.48\textwidth}
    \centering
    \includegraphics[width=\textwidth]{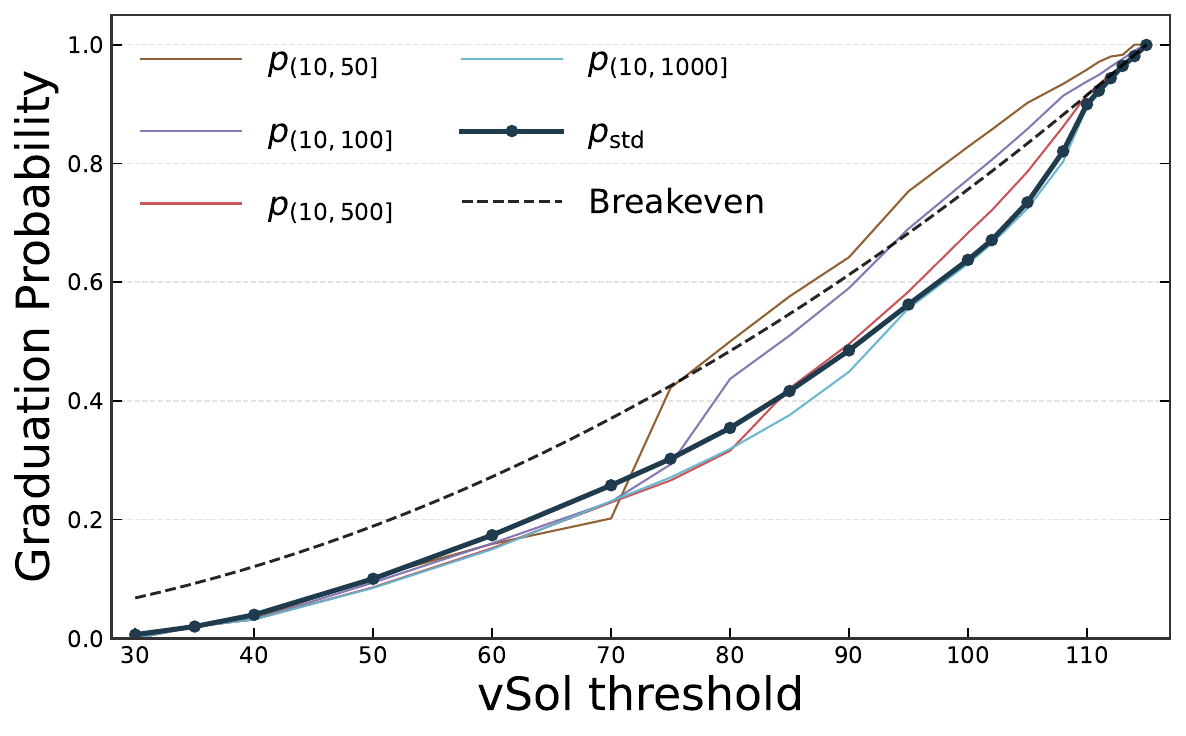}
\end{minipage}
\caption{
Graduation probability conditional on the bonding-curve state variable \textit{vSol} and on the cumulative number of swaps required to reach that level.
Curves correspond to upper bounds of 10, 50, 100, 500, and 1000 trades.
\textbf{Left panel}: full sample.
\textbf{Right panel}: same analysis after imposing a minimum of 10 trades, excluding near-instant graduation events.
%While the right panel naturally exhibits higher probabilities at low \textit{vSol}, the relative ordering of the curves is preserved, confirming that rapid liquidity accumulation robustly predicts graduation even after filtering out ultra-fast launches.responding Solana threshold in the bonding curve. 
}
\label{fig:pgrad_trades}
\end{figure}

Another relevant dimension affecting the likelihood of graduation is the intensity of trading activity prior to reaching a given level of Solana on the bonding curve.  
The same value of \textit{vSol} can be attained either through a single large transaction or via many small, possibly alternating trades.  
To disentangle these effects, we condition the graduation probability on the cumulative number of swaps observed \emph{up to the moment a given \textit{vSol} threshold is first reached}.  
This construction is strictly causal: at each point along the bonding curve, the conditioning variable depends only on past information and therefore preserves a genuinely predictive interpretation.
We consider five increasing thresholds for trading activity (10, 50, 100, 500, and 1000 swaps).  
For each threshold $N$, we estimate the probability of graduation conditional on having reached a given \textit{vSol} level with at most $N$ trades.
The left panel of Fig.~\ref{fig:pgrad_trades} displays the resulting conditional probability curves.  
A clear monotonic pattern emerges: for any fixed \textit{vSol}, tokens that reach that level with fewer trades exhibit substantially higher graduation probabilities.  
In particular, rapid accumulation of liquidity within the first few tens of transactions produces graduation probabilities that lie well above the baseline curve conditioned only on \textit{vSol}.  
Conversely, when hundreds or thousands of trades are required to reach the same amount of Solana, the graduation probability flattens and converges toward the baseline, indicating that slow and fragmented trading activity is markedly less conducive to success.

A potential concern is that the strong uplift observed at low \textit{vSol} might be mechanically driven by tokens that graduate almost instantaneously, following only a handful of very large early trades.  
Such ultra-fast events, while statistically influential, are economically uninformative and generally not exploitable in practice.

To address this issue, the right panel of Fig.~\ref{fig:pgrad_trades} repeats the same analysis after imposing a lower bound of 10 trades, thereby excluding tokens that graduate immediately.  
While this filtering naturally raises the estimated probabilities, the qualitative ordering of the curves is preserved across the entire bonding-curve range.  
Crucially, even after removing these near-instant graduations, tokens characterized by fast liquidity accumulation continue to dominate in terms of graduation likelihood.

Taken together, these results indicate that not only the amount of capital committed matters, but also the \emph{speed} at which it is deployed.  
Successful tokens are characterized by strong early momentum, where liquidity concentrates rapidly in a small number of trades.  
In contrast, prolonged accumulation through many small transactions typically signals weak collective engagement and frequently precedes stagnation.  
In this sense, liquidity velocity emerges as the single most informative predictor of graduation among all variables considered in this study.

\subsection{Conditioning on successful traders}
One of the key peculiarities of blockchain-based markets is their inherent transparency.
Because all transactions are publicly observable, it is possible to track which wallets have achieved higher profitability over time and to analyze the trading strategies associated with those addresses.
This feature enables an entire class of trading strategies based on imitation, whereby agents condition their actions on the observed behavior and past performance of other traders.
In this setting, wallet-level performance becomes an informative signal, allowing market participants to infer and potentially replicate the strategies of more successful traders.

To exploit this feature in a predictive and causally interpretable way, we split our observation window into two consecutive subperiods of two weeks each.
We use the first subperiod to identify a set of top-performing traders based exclusively on their realized profits, measured in Solana, from closing positions during that period.
Specifically, we track wallets that have accumulated the largest net amounts of SOL from selling their positions, thereby constructing an ex-ante list of traders that can be classified as ``successful'' according to past performance.
The second subperiod is then used to study whether the participation of these ex-ante identified top traders provides predictive information about the probability of graduation for newly launched tokens.
Crucially, the set of top traders is fixed before the beginning of the second subperiod and does not depend on any trading activity occurring thereafter.
This temporal separation ensures that the conditioning variable is constructed independently of future outcomes and allows for a genuinely causal interpretation in the sense of out-of-sample prediction.
Table~\ref{tab:top_wallets_comparison} reports the list of top traders identified in the first subperiod, together with the total amount of Solana received after closing their positions, the number of trades, and the breakdown between buy and sell transactions.
Interestingly, several of these wallets exhibit an extreme asymmetry in their trading behavior, having executed only sell transactions.
This pattern suggests that such wallets may act as liquidity exit points, receiving funds that were originally accumulated across multiple addresses through intermediate transfers, possibly as part of coordinated profit-taking or aggregation strategies.

\begin{table}[H]
\centering
\scriptsize
\setlength{\tabcolsep}{4pt}
\renewcommand{\arraystretch}{1.1}

\caption{Top trader wallets ranked by PnL (SOL): comparison between the full monthly dataset and the first two weeks.}
\label{tab:top_wallets_comparison}

% ===================== DATASET MENSILE =====================
\textbf{(a) Full dataset}

\vspace{0.3em}

\begin{tabular}{r l r r r r}
\hline
\textbf{Rank} & \textbf{Wallet} & \textbf{PnL (SOL)} & \textbf{\#Trades} & \textbf{\#Buy} & \textbf{\#Sell} \\
\hline
1 & 5JewENBbfKu23TLEhfXxzL4VRHwJvcLf9BCPHAuB5Rmh & 9373.141 & 1793 & 0 & 1793 \\
2 & niggerd597QYedtvjQDVHZTCCGyJrwHNm2i49dkm5zS & 3681.430 & 4028 & 596 & 3432 \\
3 & GsbjKVpmusptbt5RwwQ6z7P4cEVhqXbhRMHPwQ4hfRp4 & 3622.959 & 905 & 754 & 151 \\
4 & suqh5sHtr8HyJ7q8scBimULPkPpA557prMG47xCHQfK & 3208.460 & 19103 & 13486 & 5617 \\
5 & CeAjiFBzgNb2oBM2v3nG8u8QnmLhtS1SiRJwqVPaSwqK & 3194.397 & 632 & 0 & 632 \\
6 & 24oimYpYTiChCRHXRkNZNqw5mEE6X8RWFEWXrsVLM8PP & 2856.466 & 48 & 25 & 23 \\
7 & Dsi8ntQziuCPt16TStjG4tgviracSryHsRcH72zbwRgu & 2687.453 & 116503 & 57849 & 58654 \\
8 & 5sNnKuWKUtZkdC1eFNyqz3XHpNoCRQ1D1DfHcNHMV7gn & 2515.810 & 4723 & 1744 & 2979 \\
9 & 2ErKM3aVUjuvGhRVvYzeq6VHxqjyT9Q4P4D6xioCRZxU & 2499.163 & 520 & 0 & 520 \\
10 & 2ezv4U5HmPpkt2xLsKnw1FyyGmjFBeW7c166p99Hw2xB & 2493.568 & 66109 & 30259 & 35850 \\
\hline
\multicolumn{6}{l}{\textbf{Best wallet:} 5JewENBbfKu23TLEhfXxzL4VRHwJvcLf9BCPHAuB5Rmh} \\
\multicolumn{6}{l}{\textbf{Expected profit:} \$1{,}874{,}628.30 \,(200 \$/SOL)} \\
\hline
\end{tabular}

\vspace{1em}

% ===================== PRIME DUE SETTIMANE =====================
\textbf{(b) First two weeks of the dataset}

\vspace{0.3em}

\begin{tabular}{r l r r r r}
\hline
\textbf{Rank} & \textbf{Wallet} & \textbf{PnL (SOL)} & \textbf{\#Trades} & \textbf{\#Buy} & \textbf{\#Sell} \\
\hline
1  & GsbjKVpmusptbt5RwwQ6z7P4cEVhqXbhRMHPwQ4hfRp4 & 3017.655 & 476   & 415   & 61 \\
2  & 24oimYpYTiChCRHXRkNZNqw5mEE6X8RWFEWXrsVLM8PP & 2845.029 & 45    & 24    & 21 \\
3  & suqh5sHtr8HyJ7q8scBimULPkPpA557prMG47xCHQfK  & 2634.320 & 15220 & 10782 & 4438 \\
4  & CeAjiFBzgNb2oBM2v3nG8u8QnmLhtS1SiRJwqVPaSwqK & 2512.679 & 493   & 0     & 493 \\
5  & niggerd597QYedtvjQDVHZTCCGyJrwHNm2i49dkm5zS  & 2492.898 & 1948  & 364   & 1584 \\
6  & jECNQy2tSeA9pZpq54eVE7WWEGSeZHpeJ1sj2RsxoR1  & 2441.683 & 101   & 1     & 100 \\
7  & 5JewENBbfKu23TLEhfXxzL4VRHwJvcLf9BCPHAuB5Rmh & 2247.917 & 568   & 0     & 568 \\
8  & Ho7pNj4ABqrVzdsN34aVPyHQ27eGbWMQHhnDp6QTFHW5 & 2002.581 & 61    & 35    & 26 \\
9  & j1oAbxxiDUWvoHxEDhWE7THLjEkDQW2cSHYn2vttxTF  & 1875.038 & 5111  & 1733  & 3378 \\
10 & Dsi8ntQziuCPt16TStjG4tgviracSryHsRcH72zbwRgu & 1835.249 & 70672 & 35030 & 35642 \\
\hline
\multicolumn{6}{l}{\textbf{Best wallet:} GsbjKVpmusptbt5RwwQ6z7P4cEVhqXbhRMHPwQ4hfRp4} \\
\multicolumn{6}{l}{\textbf{Expected profit:} \$603{,}531.00 \,(200 \$/SOL)} \\
\hline
\end{tabular}

\end{table}

\begin{figure}
\centering
\includegraphics[width=0.6\textwidth]{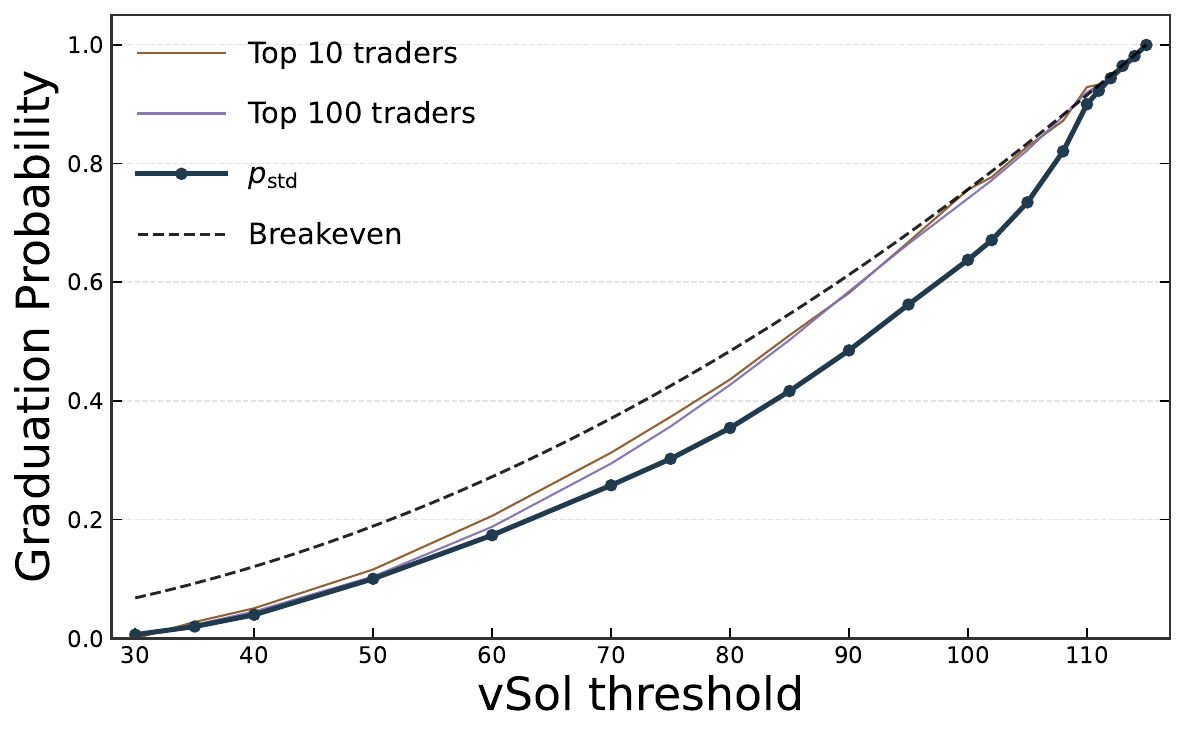}
\caption{Graduation probability conditional on \textit{vSol} and on the presence of at least one trade by a top successful trader before the corresponding Solana threshold in the bonding curve.}
\label{fig:pgrad_toptraders}
\end{figure}

In Figure~\ref{fig:pgrad_toptraders}, we condition the graduation probability on the occurrence of at least one trade executed by a trader belonging to this ex-ante defined set of top-performing wallets before a given level of \textit{vSol} is reached.
The conditioning variable is constructed in a time-consistent manner: for each value of \textit{vSol}, we only consider trades that occurred up to that point along the bonding curve and only within the second subperiod.
As a result, both the identification of top traders and the conditioning itself rely solely on information that would have been available to an observer in real time. Under this construction, the analysis admits a genuinely predictive interpretation.
The presence of a top trader should therefore be understood not as an ex-post correlate of success, but as an ex-ante signal reflecting the early involvement of agents who have demonstrated superior performance in the recent past.

The resulting conditional graduation-probability curve lies above the baseline probability conditioned only on \textit{vSol}, indicating that the early participation of ex-ante identified top traders is associated with a higher likelihood of graduation. This indicates that some traders are able to take advantage of the this graduation mechanism and, therefore, they insert information in the market. 

Note that, however, the two curves still are below the breakeven curve. Meaning that using a simple buy and hold strategy gaining would be unlikely. This is surely due, in part, to the simplicity of the strategy. However, it is also important to consider that the effect of top traders usually reflects two opposing mechanisms.
On the one hand, early entry by experienced traders may accelerate initial liquidity accumulation and help coordinate early demand.
On the other hand, these same traders tend to exit positions rapidly, often around salient events such as graduation, thereby limiting the sustained buying pressure required to complete the launchpad phase.
Overall, the presence of top traders emerges as a double-edged signal: it conveys information about informed participation, yet it may also coincide with opportunistic behavior that constrains the long-run stability of the bonding curve.

\subsection{Conditioning with respect top token creators}

Another important dimension for understanding the lifecycle of newly issued tokens concerns the identity and behavior of their creators.
Token creators represent a relatively new class of market participants that has emerged with the advent of low-cost, high-throughput launchpad platforms such as Pump.fun.
These platforms drastically reduce the cost of token creation, enabling a single address to deploy a very large number of tokens within a short time span.
Creators do not merely influence the market by increasing the supply of new assets.
They can also actively shape the subsequent evolution of the tokens they issue by participating directly in trading activity after creation.
In particular, creators enjoy a structural advantage in timing, as they are able to interact with the token immediately upon deployment and can therefore be among the earliest buyers.
This early access potentially allows creators to influence price dynamics during the most fragile phase of the token’s lifecycle, when liquidity is still thin and coordination effects are strongest.

At the same time, the design of Pump.fun imposes important constraints on creator behavior.
Unlike many decentralized exchange environments, creators cannot directly extract value through liquidity-removal mechanisms such as rug pulls.
All liquidity is managed by the protocol, and creators have no privileged control over the pool.
As a result, creators can affect the market only through standard trading actions, such as buying and selling the token along the bonding curve.
This restriction isolates a well-defined class of strategic behaviors, most notably pump-and-dump strategies, that operate purely through trading rather than through contract-level manipulation. These features make creator activity particularly informative for studying early token dynamics.

\begin{figure}[H]
\centering
\includegraphics[width=0.6\textwidth]{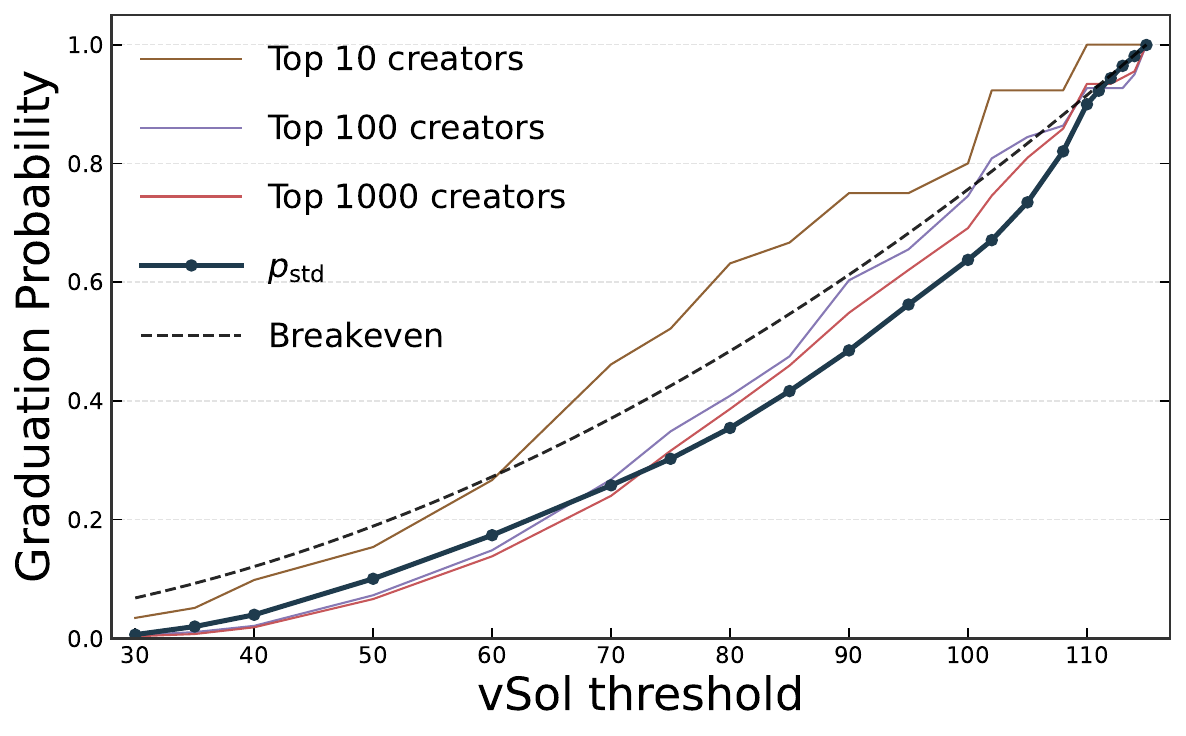}
\caption{}
\label{fig:prob_creator}
\end{figure}
As we did in the last section with successful traders, we investigate the predictive information contained in successful token creators. Table \ref{tab:top_creators_ratio} reports the top 10 token creators in the month of September, some of whom issued more than 2,000 tokens within a single month. This role is further incentivized by the platform, contributing to its dominance among existing launchpads, as it grants token creators a share of the fees generated by swaps involving their own tokens. Specifically, creators receive a 0.3\% fee during the virtual AMM phase, and a dynamic fee in the real \textit{PumpSwap} AMM phase that ranges from 0.950\% down to 0.050\% as the amount of SOL locked in the bonding curve continues to increase.

\begin{table}[H]
\centering
\scriptsize
\setlength{\tabcolsep}{4pt}
\renewcommand{\arraystretch}{1.1}

\begin{tabular}{r l r r r}
\hline
\textbf{Rank} 
& \textbf{Creator} 
& \textbf{Ratio} 
& \textbf{$N_{\rm grad}$} 
& \textbf{$N_{\rm tot}$} \\
\hline
1  & CeZbzsqje88U7GCCrAiGUr5BrZrBJ3AkQMyG9aggDiqQ & 0.084 & 9  & 107 \\
2  & HyRzY2SNzGSFvhJkCepj6DRtR3LaHCSyNKJ4HMYktbU4 & 0.058 & 3  & 52  \\
3  & ESBjYRXRkYUCqoY54V6AB3BXvNW4P37bBhgutfPejJ2X & 0.050 & 3  & 60  \\
4  & Hh3zeEWtMLM24mQBnomf7Fviem5xEMwJGMxWchhX4gwK & 0.041 & 3  & 73  \\
5  & GH9yk8vgFvHnAD8JZqXxr3hBN1Lr1mJ9NPzrP5mVqiJe & 0.041 & 11 & 268 \\
6  & Gak3gXmn3N6JUTggPanQA52yhwzLJ9fdM3SamZ42N5HD & 0.039 & 2  & 51  \\
7  & 8deJ9xeUvXSJwicYptA9mHsU2rN2pDx37KWzkDkEXhU6 & 0.036 & 2  & 56  \\
8  & GFA2t9fZVY3jC7wDBPDB8FdsVNzLWgK44J4q7XpP476a & 0.031 & 3  & 97  \\
9  & 5nNvKhVy27NLYq4xVcCV84QY7v19RQQxYQEfHz7oP1QZ & 0.027 & 7  & 263 \\
10 & C9TGKPhBf3ovo4RrjeP2bThyhP37DHLgkTqkkvFE4mMs & 0.023 & 2  & 88  \\
\hline
\end{tabular}

\caption{Top creator ranking under filters: total tokens $\ge 50$, graduation step $\ge 10$, and ratio $>0.001$.}
\label{tab:top_creators_ratio}
\end{table}

\begin{table}[ht]
\centering
\small
\setlength{\tabcolsep}{6pt}
\renewcommand{\arraystretch}{1.25}
\begin{tabular}{
c|
c|c|
c|c|
c|c|
c|c|
c|c
}
\toprule
\textbf{${\rm vSol_{min}}>\tau$}
& \multicolumn{2}{c|}{\textbf{Std}} 
& \multicolumn{2}{c|}{\textbf{isBot$>0.3$}} 
& \multicolumn{2}{c|}{\textbf{Trades$\ge 10$}} 
& \multicolumn{2}{c|}{\textbf{Top100 wallets}} 
& \multicolumn{2}{c}{\textbf{Top100 creators}} \\
\midrule
\textbf{$\tau$}
& {$N_{\text{elig}}$} & {$p$}
& {$N_{\text{elig}}$} & {$p$}
& {$N_{\text{elig}}$} & {$p$}
& {$N_{\text{elig}}$} & {$p$}
& {$N_{\text{elig}}$} & {$p$} \\
\midrule

30  & 652768 & 0.006 & 372960 & 0.010 & 0     & 0.000 & 3885 & 0.004 & 7183 & 0.005 \\
50  & 40817  & 0.100 & 32794  & 0.112 & 36999 & 0.085 & 8968 & 0.104 & 523  & 0.073 \\
80  & 11557  & 0.354 & 8181   & 0.436 & 11163 & 0.326 & 2679 & 0.427 & 93   & 0.409 \\
100 & 6417   & 0.638 & 4611   & 0.748 & 6355  & 0.616 & 1649 & 0.741 & 51   & 0.745 \\
\bottomrule
\end{tabular}

\vspace{6pt}
\footnotesize
\textit{Note:} Top100 wallets/creators are computed on the last two weeks with the top wallets/creators defined on the first two weeks of the month. 
\end{table}

Table~\ref{tab:top_creators_ratio} reports the top token creators identified during the first two weeks of September.
We define a creator as \emph{successful} if the fraction of their tokens that reach graduation is at least \(0.001\), subject to two additional constraints.
First, each token must have recorded at least 10 trades during its lifetime, in order to exclude cases in which a creator forces an almost immediate graduation through isolated transactions.
Second, the creator must have issued at least 50 tokens in total.
These thresholds are chosen to ensure a sufficiently large and statistically meaningful sample of tokens per creator.

Successful token creators are identified using data from the \emph{first half} of the month only.
Following the same logic adopted for top traders, we then perform an out-of-sample analysis by computing, over the subsequent two weeks, the probability of graduation conditional on the token being created by one of these selected creators.
This procedure ensures that the conditioning does not exploit future information and preserves a predictive interpretation.

We repeat the analysis by considering different sets of creators, namely the top 10, top 100, and top 1000 successful creators.
The resulting conditional graduation probabilities are shown in Figure~\ref{fig:prob_creator}.
We observe that, as the bonding-curve threshold \textit{vSol} increases, the probability of graduation rises sharply.
In the case of the top 10 creators, the conditional probability even exceeds the economic breakeven curve introduced earlier, indicating that creator identity can carry economically relevant predictive information at sufficiently advanced stages of the bonding curve.

It is important to stress, however, that the statistical support in this conditioning is limited.
As reported in Table~\ref{tab:top_creators_ratio}, the number of eligible tokens created by these highly selective groups of creators is relatively small, especially for the top 10 case.
As a result, while the effect is suggestive and consistent with a role of experienced creators in sustaining tokens that have already accumulated liquidity, the corresponding estimates should be interpreted with appropriate caution.

\section{Pump \& Dump of the tokens}

Pump-and-dump behavior is a well-known feature of low-friction, low-screening token issuance environments: early participants can accumulate inventory during the initial trading phase and subsequently sell it in a concentrated manner once sufficient liquidity has formed \cite{Bello2023LLD, Golmohammadi2014Detecting}. In our setting, we deliberately concentrate on \emph{dumps} rather than \emph{pumps} \cite{Hamrick2021PumpAndDumpEcosystem, Kamps2018ToTheMoon}. In practice, the buy-side leg is intrinsically hard to detect comprehensively: accumulation can be spread over time and fragmented across multiple addresses that are ultimately controlled by the same trader, making coordinated buying largely indistinguishable from ordinary, dispersed demand in on-chain data. By contrast, the exit leg is constrained by execution incentives. To maximize monetization before liquidity conditions deteriorate, a trader typically needs to unwind inventory rapidly and in a highly concentrated manner, using one or a small set of wallets. This concentration makes dump episodes both economically more relevant and empirically more identifiable.

\subsection{Shewhart-type rule for dump detection}
\label{subsec:shewhart_dump}

In our setting, we operationalize a \emph{dump} as a statistically significant negative price shock detected via the rule introduced in this subsection. Motivated by the abrupt drawdowns in Figs.~\ref{fig:crash_drop}--\ref{fig:crash_volume}, we implement a transparent control-chart procedure to flag dump events during the bonding-curve phase \cite{Fantazzini2023DetectingCryptoPD}. For each token $i$, let $p_{i,t}$ denote the marginal price implied by the curve at observation time $t$ (equivalently, the on-curve spot price). We work with log-returns
\begin{align}
    r_{i,t} \equiv \log \frac{p_{i,t}}{p_{i,t-1}} \,\,.
\end{align}

A Shewhart chart requires an ``in-control'' baseline \cite{Champ1987ShewhartRunsRules}. For each token we estimate a reference location and dispersion using an initial window that precedes the first major run-up. To ensure that the robust scale estimator is well-defined and statistically stable, we restrict attention to tokens for which the baseline window contains at least $30$ to a maximum look-back of $200$ trades  (i.e., at least $30$ observations of $r_{i,t}$). This minimum-sample requirement provides a meaningful empirical distribution from which to estimate dispersion and avoids spurious ``extreme'' events driven by very short or illiquid histories.

To mitigate the impact of heavy tails and  microstructure noise, we replace the mean--standard-deviation pair with a median--MAD specification. Specifically, let
\begin{align}
    m_i \equiv \mathrm{median}\!\left(r_{i,t}\right),
    \qquad
    \mathrm{MAD}_i \equiv \mathrm{median}\!\left(\left|r_{i,t}-m_i\right|\right),
\end{align}
computed over the baseline observations. We then define the Gaussian-consistent MAD scale as
\begin{align}
    \sigma_{\mathrm{MAD},i} \equiv \frac{1}{0.67449}\,\mathrm{MAD}_i,
\end{align}
where the factor $1/0.67449 = \sqrt{2}~ {\rm erf}^{-1} \frac{1}{2}$ ensures consistency with the standard deviation under a gaussian reference model \cite{RousseeuwCroux1993Alternatives}.

We declare a dump signal at the first time $t$ such that the log-return violates a robust $k$-sigma lower control limit,
\begin{align}
    \label{eq:shewhart_rule}
    \tau_i^{D} \equiv \inf\left\{t:\ r_{i,t} < -k\,\sigma_{\mathrm{MAD},i}\right\},
    \qquad k=4.
\end{align}

\subsection{Prevalence and stylized facts}
\label{subsec:dump_stylized_facts}

\begin{figure}[t]
    \centering
    \includegraphics[width=0.7\textwidth]{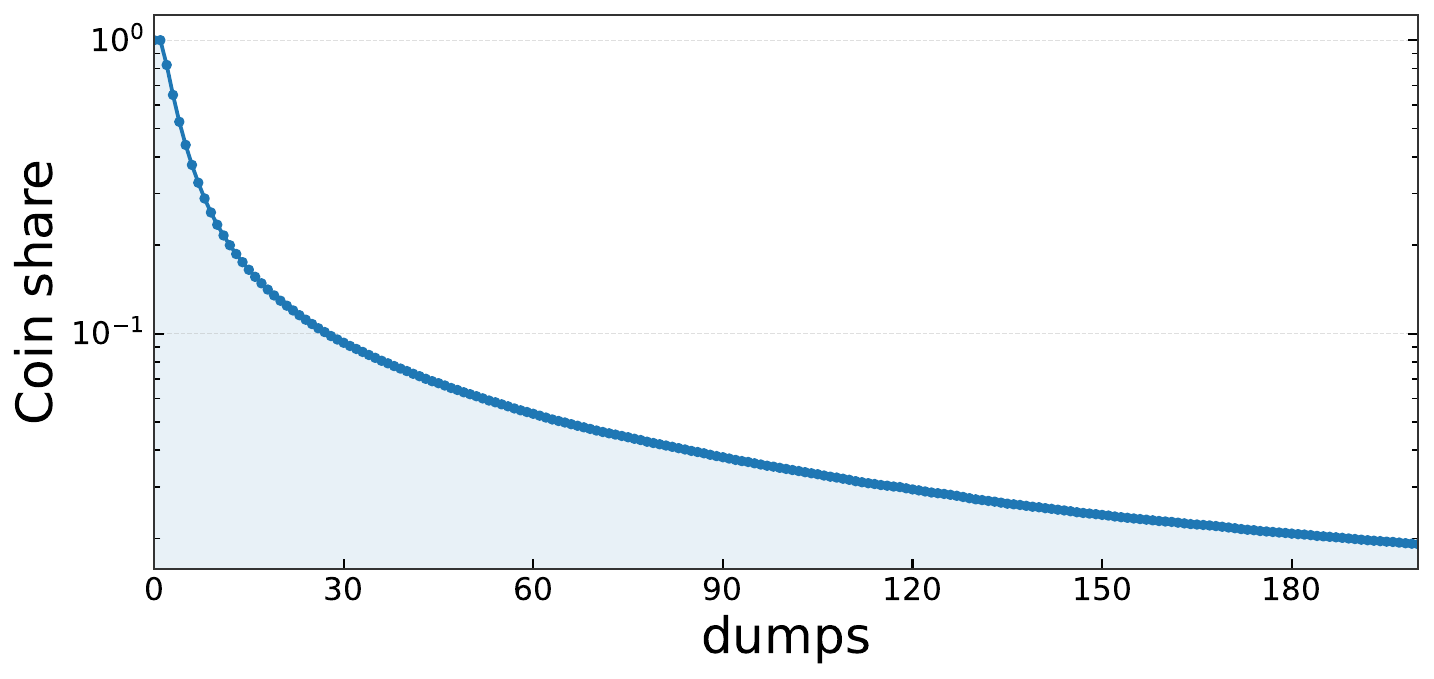}
    \caption{\textbf{Coverage curve.} For each dump threshold $k$, the curve reports the fraction of coins such that $\texttt{dumps} \ge k$. The tail of the distribution is preserved by appending a final point at $k = \max_k$, which aggregates all observations with $\texttt{dumps} \ge \max_k$.}
    \label{fig:dump_count_dist}
\end{figure}

Under the definition above, dumps are not a marginal phenomenon: out of $184{,}282$ tokens in our sample that had at least $30$ swaps, $169{,}938$ (92.22\%) exhibit at least one dump event. 
Among the $184{,}282$, only $2.55\%$ graduate. Figure~\ref{fig:dump_count_dist} reports the distribution of tokens by the number of detected dumps, highlighting a strongly right-skewed structure: most tokens experience no dump or only a small number of events, while a minority displays repeated episodes.

\begin{figure}
    \includegraphics[width=0.7\textwidth]{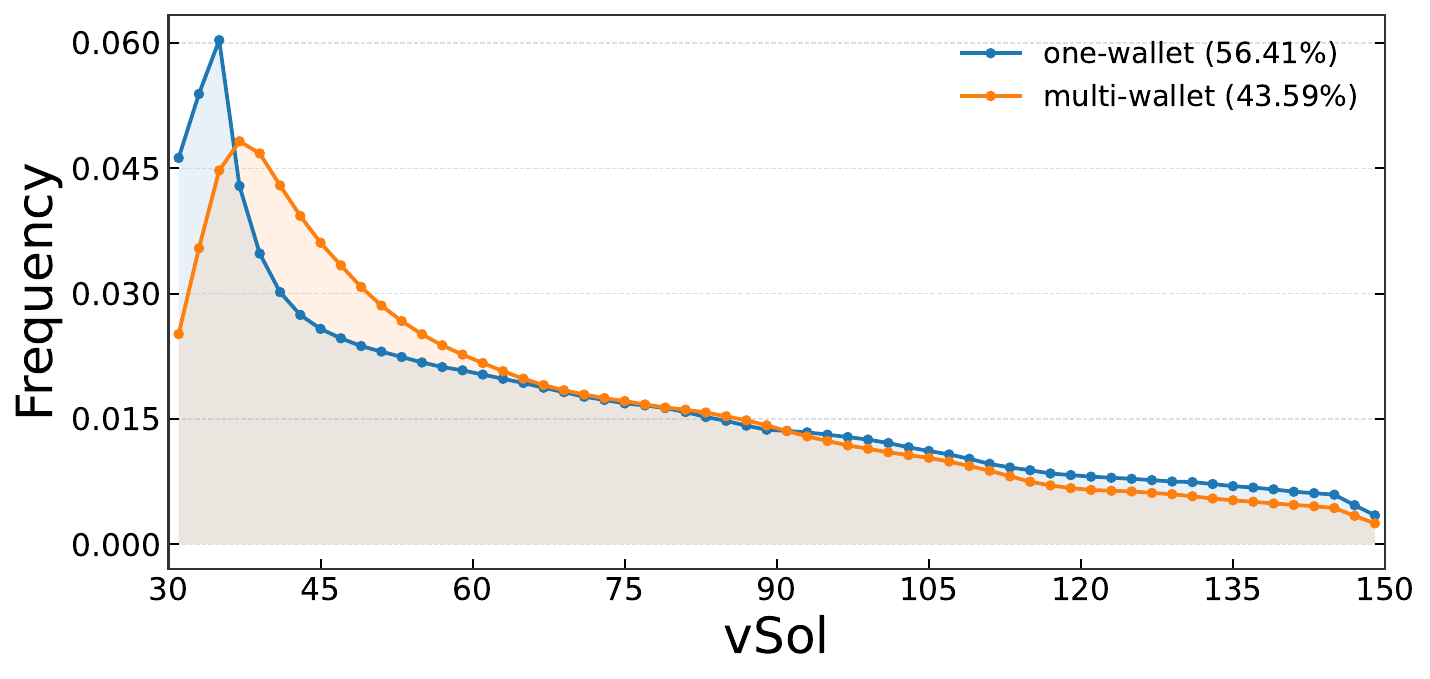}
	   	\caption{Distribution of the value of SOL in the bonding curve when the dump occurs, shown separately for (i) one-wallet dump episodes and (ii) for multi-wallet dump episodes.}
		\label{fig:crash_vsol}
\end{figure}

\begin{figure}
    \includegraphics[width=0.7\textwidth]{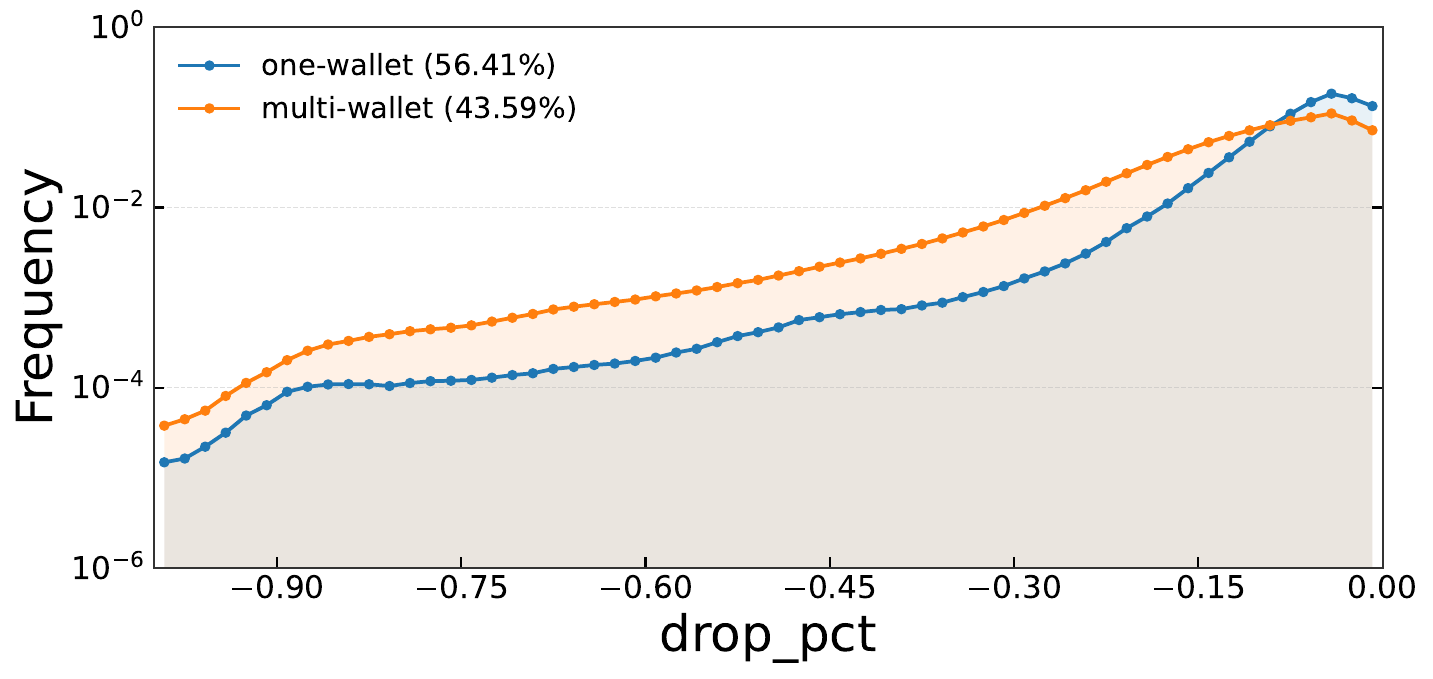}
	   	\caption{Distribution of drop percentage in a dump event, shown separately (i) in one-wallet dump episodes and (ii) in multi-wallet dump episodes.}
		\label{fig:crash_drop}
\end{figure}

\begin{figure}
    \includegraphics[width=0.7\textwidth]{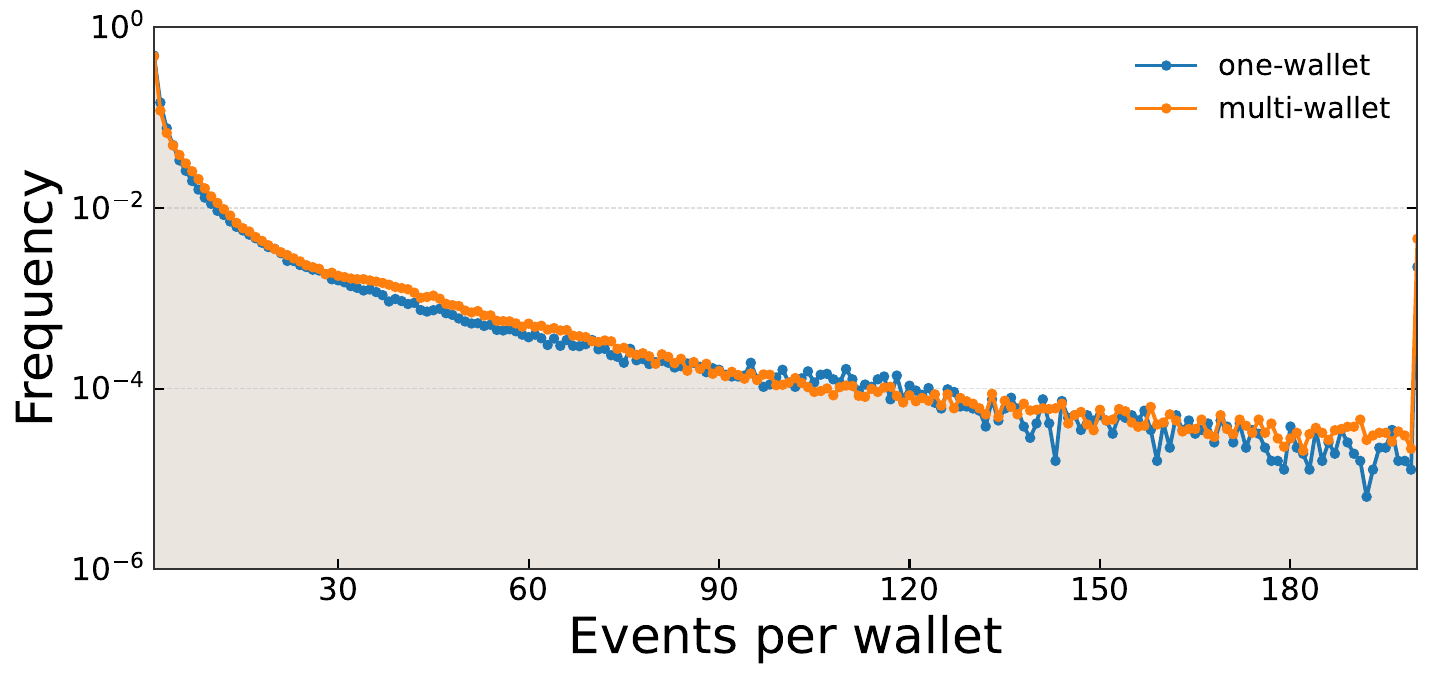}
	   	\caption{Distribution of the number of dump events executed by each wallet, shown separately for (i) wallets involved in one-wallet dump episodes and (ii) wallets participating in multi-wallet dump episodes. For each $\texttt{Events per wallet}$, the histogram reports the frequency of wallets that performed exactly $\texttt{Events per wallet}$ dumps.}
		\label{fig:crash_volume}
\end{figure}

From Figs.~\ref{fig:crash_vsol}--\ref{fig:crash_volume} we observe a recurring pattern that is consistent with profit extraction by token creators and, more generally, with pump-and-dump \cite{Balcilar2023RiskSpillover, Yu2025ImprovingCryptoPDD}. In particular, Fig.~\ref{fig:crash_vsol} shows that large drops tend to cluster at specific levels of $vSol$, i.e., when a non-negligible amount of SOL has already accumulated along the curve, while Figs.~\ref{fig:crash_drop} and \ref{fig:crash_volume} jointly characterize the severity of such events (drop magnitudes) and the associated concentration of selling activity across wallets (single- vs.\ multi-wallet episodes). A natural interpretation is that, once the pool has accumulated enough SOL to make the liquidation economically profitable (often prior to graduation), the creator (or a tightly connected set of early wallets) offloads a substantial inventory of tokens acquired since launch and cashes out in SOL. The timing is not accidental: in the late stages of the bonding curve, the price becomes mechanically more sensitive to net order flow, so a concentrated sell wave can trigger an abrupt drawdown.

The incentive to exit before graduation is further strengthened by a liquidity discontinuity at the transition. During the launchpad phase, pricing is produced by a bonding-curve state that includes virtual reserves. At graduation the market moves to a post-migration AMM supported only by real reserves. Even if the marginal price is approximately continuous at the boundary by construction, the depth is not: the disappearance of virtual liquidity reduces the effective inventory against which sells are absorbed. This effect is visible in the expected proceeds of a sale executed just before versus just after graduation.

Fix a token inventory $\Delta y$ held by a trader. Consider first a sale \emph{before} graduation, when the bonding-curve state is approximately $x = 115$~SOL and $y = 0.2799\cdot 10^{9}$ tokens. Neglecting fees, the constant-product swap gives
\begin{align}
    \label{eq:liq_drop_before}
    \Delta x
    = x - \frac{xy}{y+\Delta y}
    = \frac{x\Delta y}{y+\Delta y}
    = P\,\frac{\Delta y}{1 + \frac{\Delta y}{y}},
\end{align}
where $P \equiv \frac{x}{y}$ denotes the marginal price before the swap.

Consider next the same trade executed \emph{after} graduation, when the corresponding post-migration AMM state is approximately $x' = 85$~SOL and $y' = 0.2069\cdot 10^{9}$ tokens. The SOL received is
\begin{align}
    \label{eq:liq_drop_after}
    \Delta x'
    = \frac{x'\Delta y}{y' + \Delta y}
    = P\,\frac{\Delta y}{1 + \frac{\Delta y}{y'}},
\end{align}
where the equality uses the fact that the boundary is calibrated so that\footnote{Remember that continuity of the marginal price does not imply continuity of depth.} $\frac{x'}{y'} \simeq \frac{x}{y}=P$. Since $y>y'$, it follows that $\frac{\Delta y}{y} < \frac{\Delta y}{y'}$ and therefore
\begin{align}
    \label{eq:liq_drop}
    \frac{\Delta x}{\Delta x'}
    = \frac{1 + \frac{\Delta y}{y'}}{1 + \frac{\Delta y}{y}}
    > 1
    \qquad \Rightarrow \qquad
    \Delta x > \Delta x'
    \ \ \ \forall\,\Delta y>0.
\end{align}
Hence, holding the same token inventory, selling strictly before graduation is more profitable than selling immediately after graduation, providing a simple mechanical rationale for the pre-graduation sell pressure documented in Figs.~\ref{fig:crash_drop} and \ref{fig:crash_volume}. 

These results link directly to graduation prediction. Because selling strictly before graduation yields higher proceeds, early actors have a clear incentive to dump while the token is still on the bonding curve. The high incidence of pre-graduation dumps therefore acts as a frequent mechanism that cuts the trajectory short: once a dump occurs, the token is far less likely to keep attracting the net inflows required to reach the graduation threshold. In this sense, repeated dump activity makes the virtual bonding-curve market highly manipulable, since coordinated exit behavior can deterministically depress (or terminate) the graduation path.

Predicting the pump leg is considerably harder. As discussed above, accumulation is often executed through multiple wallets, and it is common for the subsequent dump to be carried out by a single wallet after the pump has been distributed across several addresses, with tokens consolidated into the selling wallet at a later time. Identifying such patterns reliably requires deeper wallet-linkage and transfer-graph analysis. For this reason, we leave a systematic study of pump-prediction mechanisms to future work.

\section{Conclusions}
%\subsection{Number of distinct wallets trading the coin}
This study provides an empirical characterization of token launches on Pump.fun and of the conditions under which newly created tokens reach the protocol-defined \emph{graduation} threshold. The economic motivation for this setting is that tokens are often launched as a financing and coordination primitive for projects with substantive underlying development goals: they are used to bootstrap ecosystems, attract early communities, and establish an initial market valuation under transparent, on-chain rules. Using a fully on-chain dataset reconstructed from Solana transaction logs over a one-month window (September 2025), we document an ecosystem operating at very large scale, with hundreds of thousands of new tokens and a graduation rate well below one percent. This extreme class imbalance motivates a probabilistic framing: rather than attempting to assess project fundamentals directly, we treat graduation as a binary, protocol-level outcome and quantify how its likelihood evolves as a function of the bonding-curve state, thereby offering a scalable screening signal for early traction among project-backed launches.

Our baseline quantity is the empirical probability of graduation conditional on the amount of SOL accumulated in the bonding curve. As expected, this probability increases monotonically with \textit{vSol} and approaches one near the deterministic threshold. However, conditioning on \textit{vSol} alone leaves unexplained substantial heterogeneity. We therefore augment the conditioning set with variables that proxy market microstructure and participant composition, constructed in a way that is observable up to the current point on the curve. The resulting conditional curves reveal that not all paths to the same \textit{vSol} are equivalent: rapid accumulation of liquidity, achieved in a small number of trades, is systematically associated with higher graduation likelihood. For tokens associated with substantive projects, this pattern is consistent with the interpretation that early, coordinated demand reflects stronger initial conviction and faster community formation, which in turn sustains the net inflow required to reach migration.

A second set of results concerns the role of agent types. Tokens whose early activity is dominated by transactions classified as bot-like exhibit a lower probability of graduation beyond intermediate \textit{vSol} levels, suggesting that algorithmic participation is often correlated with short-horizon trading that does not support sustained capital formation. The presence of historically profitable (“successful”) traders provides at most a modest uplift in graduation probability and should be interpreted with caution, since the identification of such traders can embed ex-post information. In contrast, conditioning on prolific token creators does not improve predictive performance relative to the baseline and is associated with patterns consistent with creator-specific incentives that may not align with long-horizon project value creation.

Several extensions follow naturally. First, the predictive framework could be cast as a hazard-rate (survival) model, so graduation likelihood can be updated continuously over time and along the curve, while properly handling right-censoring. Second, bot detection and the definition of “successful traders” could be made more robust by using out-of-sample evaluation windows and richer behavioral features, reducing look-ahead bias. Third, the profit-extraction channel deserves dedicated accounting of creator-level PnL, including cross-wallet transfers and coordinated selling, to measure how value is redistributed and how these flows interact with project-funding objectives. More broadly, our results suggest that, in bonding-curve launchpads, the composition and pace of early trading carry material predictive signal for protocol-level success and early project traction, while also highlighting structural incentives that can degrade post-graduation market quality if not carefully managed. Finally, future work could expand the feature set with token metadata and platform-level signals, e.g. token name and image, or web-scraped indicators from Pump.fun top-token lists and other on-site rankings.

\subsection*{Acknowledgements}
    FL and FT acknowledge support from the grant PRIN2022 DD N. 104 of February 2, 2022 ”Liquidity and systemic risks in centralized and decentralized markets”, codice proposta 20227TCX5W - CUP J53D23004130006 funded by the European Union NextGenerationEU through the Piano Nazionale di Ripresa e Resilienza (PNRR).\\

\bibliography{biblio}

%%%%%%%%%%%%%%%%%%%%%%%%%%%%%%%%%%%%%%%%%%%%%%%%%%%%%%%%%%%%%%%%%%%%%%%%%%%%%%%%

%%%%%%%%%%%%%%%%%%%%%%%%%%%%%%%%%%%%%%%%%%%%%%%%%%%%%%%%%%%%%%%%%%%%%%%%%%%%%%%

%%%%%%%%%%%%%%%%%%%%%%%%%%%%%%%%%%%%%%%%%%%%%%%%%%%%%%%%%%%%%%%%%%%%%%%%%%%%%%%%

\end{document}